\documentclass[preprint,12pt]{elsarticle}

\usepackage{amssymb}
\usepackage{amsmath}
\usepackage{textcomp}

\newcommand{\Cell}[2]{\mbox{\textcolonmonetary}_{#1} \vert V_{#2}}
\newcommand{\Cellonly}[1]{\mbox{\textcolonmonetary}_{#1}}

\journal{Journal of Computational Physics}

\begin{document}

\begin{frontmatter}

\title{A soft particle dynamics method based on shape degrees of freedom}


\author[INRAE,UM]{Yohann Trivino}
\author[UGA]{Vincent Richefeu}
\author[UM]{Farhang Radjai}
\author[INRAE]{Komlanvi Lampoh}
\author[INRAE]{Jean-Yves Delenne}

\affiliation[UM]{organization={LMGC, CNRS, University of Montpellier},
            addressline={34090}, 
            city={Montpellier},
            country={France}}

\affiliation[INRAE]{organization={IATE, INRAE},
            addressline={Institut Agro, University of Montpellier, 34000}, 
            city={Montpellier},
            country={France}}

\affiliation[UGA]{organization={Laboratoire 3SR, CNRS, University Grenoble Alpes},
            addressline={38400}, 
            city={Grenoble},
            country={France}}


\begin{abstract}
In this paper, we present a 2D numerical model developed to simulate the 
dynamics of soft, deformable particles. To accommodate significant particle deformations, 
the particle surface is represented as a narrow shell composed of mass points that interact 
through elasto-plastic force laws governing their linear and angular relative 
displacements. Particle shape changes are controlled by these interactions, 
in conjunction with a uniform particle core stiffness. We calibrate and verify this model by 
comparing the deformation of constrained beams under load with 
theoretical predictions. Subsequently, we explore the diametral compression of 
a single particle between two walls, focusing on the influence of the particle core 
stiffness and shell plasticity. Our findings indicate that increased core stiffness 
reduces particle volume change and promotes the development of faceting through 
flat contact areas with the walls.
To further illustrate the model's capabilities, we apply it to the uniaxial 
compaction of a granular material composed of core-shell particles. 
We show that, depending on the core stiffness and shell plasticity, 
the compaction leads to either a significant reduction of particle volumes 
or an improved pore filling due to particle shape changes. 
At high compaction, particle shapes vary: elastic particles 
without core stiffness become mostly elongated, elastic particles with core  
stiffness form polygonal shapes, while plastic particles develop elliptical or 
highly irregular forms. Finally, we simulate the tensile fracture of a tissue composed 
of elastic or plastic cells, illustrating the model’s potential applicability to 
soft tissues that undergo both large cell deformations and fracture.
\end{abstract}



\begin{keyword}
Soft particles \sep Core-shell particles \sep Granular materials 
\sep Cellular materials \sep Discrete Element Method 
\sep Particle shape \sep Compaction \sep Fracture
\end{keyword}

\end{frontmatter}


\section{Introduction}
\label{sec:intro}

The study of soft particles is a diverse and rapidly evolving field within 
materials science and the physics of complex systems. 
These soft particles encompass a wide range of materials, including 
hydrogels \cite{Heyes2009}, polymers \cite{Bonnecaze2010,tang2021liquid}, 
colloids \cite{Cloitre2003,Likos2006,deshmukh2015hard}, 
emulsions \cite{pineres2022environmentally}, 
plant cells \cite{Coen2024,Daher2015}, and red blood cells \cite{Yazdani2021}. 
Understanding the mechanical behavior of these materials has broad implications 
across multiple disciplines. For instance, deformable polymer particles 
can be engineered to create innovative materials with customized 
properties \cite{zhu2014photoreconfigurable,weerakoon2017thermoplastic,gora2011melt}. 
Similarly, colloids, known for forming complex structures in response to subtle 
interactions, are used in the design of more efficient photovoltaic 
materials and in the development of controlled-release drug delivery systems. 
The fields of medicine and cosmetics also benefit from research 
into soft particles \cite{bolzinger2011nanoparticles}. For example, by studying 
how emulsions interact with biological cells, researchers can develop more 
effective and better-targeted pharmaceutical formulations. 

Plant cells and red blood cells belong to the broader class of 
core-shell particles, which consist of solid, liquid, or gas cores 
surrounded by solid shells  \cite{galogahi2021thermal,
galogahi2020core,yadav2023core,ghosh2012core}. 
The combined effects of their core 
and shell materials make  them suitable for many 
applications such as drug delivery, chromatography, catalysis, and design of artificial cells. 
The core-shell particles are assembled from inorganic, organic, 
polymeric, or crystalline materials with sizes ranging from nanometer to  
micrometer. Hollow nanoparticles are of particular interest because 
of their lower density compared to their solid counterparts, and 
they have been explored for their use in catalysis. 
The mechanical properties of core-shell particles are essential for their 
performance in applications. In exception to a few studies \cite{ghosh2012core,fery2007mechanical,huang2021models}, 
most past studies have been concerned with the fabrication and initial 
biological characterization of core-shell particles rather than their 
mechanical behavior. 

What connects these materials, despite their diverse physical natures, 
is their common ability to undergo significant deformation due to both 
collective particle rearrangements and changes in particle volume or 
shape under external and internal stresses. These characteristics profoundly 
alter the space-filling properties of soft-particle systems compared to 
hard particle packings, where the packing fraction cannot exceed the random 
close packing (RCP) limit \cite{Berryman1983,Torquato2000}. 
The rheological behavior of highly deformable particle systems confined 
beyond this jamming limit has only recently been 
studied systematically \cite{Nezamabadi2015,Boromand2018,cheng2022hopper}.

The standard method for simulating granular materials is the Discrete 
Element Method (DEM) \cite{Radjai2011}. This technique consists in integrating 
the equations of motion for the translational and rotational degrees 
of freedom of each particle, accounting for their contact interactions, 
and employing a time-stepping scheme. The primary assumption of DEM is 
that particles are undeformable and lack shape degrees of freedom, 
though small overlaps are permitted to represent contact elastic or plastic 
deflections between particles. Consequently, DEM is unsuitable for simulating 
soft particles, which require shape degrees of freedom.
Several numerical models have been used for deformable 
particles. The Finite Element Method (FEM) and various mesh-free methods 
address particle deformations and frictional contact interactions by solving 
the governing equations \cite{barkanov2001introduction,alberty2002matlab,Guener2015,Mollon2016,CardenasBarrantes2022}.  The Material Point Method (MPM) combines aspects of both Lagrangian 
and Eulerian methods, allowing for efficient handling of large deformations 
and particle interactions \cite{Nezamabadi2015,Nezamabadi2024}. 
Although these methods provide high precision and accuracy in modeling 
deformable particles, they are often numerically inefficient, 
particularly regarding computational cost and implementation complexity.

Efficient numerical methods incorporating shape degrees of freedom 
within the DEM have been proposed. For instance, Mollon 
developed a Soft Discrete Element Method, where rotations and center-of-mass 
displacements are managed similarly to DEM, but elliptical particles can 
adjust their aspect ratio through stepwise variations in strain (assumed to be uniform) 
in response to changes in average stress due to contact forces \cite{mollon2022soft}. 
However, this model is limited to elliptical particles and cannot be used to 
simulate arbitrary particle shapes. 
Boromand et al. introduced a versatile Deformable Particle (DP) method, 
where the degrees of freedom are represented by mass points (or vertices) 
on the particle's boundary \cite{Boromand2018}. The evolution of these 
boundary points is governed by a shape-energy function, 
which is minimized with each increment of imposed displacements to 
determine the new positions of the points. 
This approach allows for arbitrary deformations from a reference shape 
defined by the initial positions of the vertices. The DP model has been 
applied to study the jamming of soft grains \cite{Boromand2018} as well as 
the isotropic compaction and hopper flow of deformable 
particles \cite{cheng2022hopper,Boromand2019}. 
From the viewpoint of mechanical behavior, the DP  
model can be considered as a method for the simulation of 
hollow-core thin-shell particles rather than solid particles.  

In this paper, we present a 2D numerical method that builds on the 
principles of the DEM by incorporating frictional contact interactions 
between individual particles, but with the added capability of handling 
large deformations. In close analogy with the DP approach, 
our model introduces shape degrees of freedom 
by representing the particle surface as a shell composed of mass points interacting 
through elasto-plastic force laws. To better control particle volume changes, 
the model incorporates a core stiffness in addition to the stretch and bending 
stiffness parameters associated with the shell elements. 
The Soft Particle Dynamics (SPD) method proposed here is 
fully dynamic and can simulate both elastic 
and plastic deformations of particles. We shall see that  
the mechanical behavior of a single particle can vary significantly based 
on the chosen stiffness parameters and shell plasticity. 
We highlight the versatility of this method by applying it to the 
compaction of deformable particles and the fracture of cellular tissues.

Section \ref{sec:core-model} presents a detailed description of the approach. 
Section \ref{sec:verif} covers the calibration of the method, 
focusing on its application to beams subjected to point forces. 
In Section \ref{sec:compression}, we examine how model parameters 
affect the mechanical response of a single particle under diametral 
compression between two flat walls. 
Section \ref{sec:compaction} discusses the compaction of an 
assembly of deformable particles. For demonstration purposes, Section \ref{sec:fracture} 
presents two examples of cellular tissue fracture under 
tensile deformation. Finally, Section \ref{sec:conclusion} concludes with 
a summary of the key findings and potential future directions for this work. 

\section{Model description}
\label{sec:core-model}

In this section, we outline the main ingredients of the model 
and introduce the notations that will be used throughout the paper. 
It is important to note that both the model and the examples discussed 
are in two dimensions under plane-stress conditions. 
  
\begin{figure}[tbh]
\centering
\includegraphics[width=0.4\columnwidth]{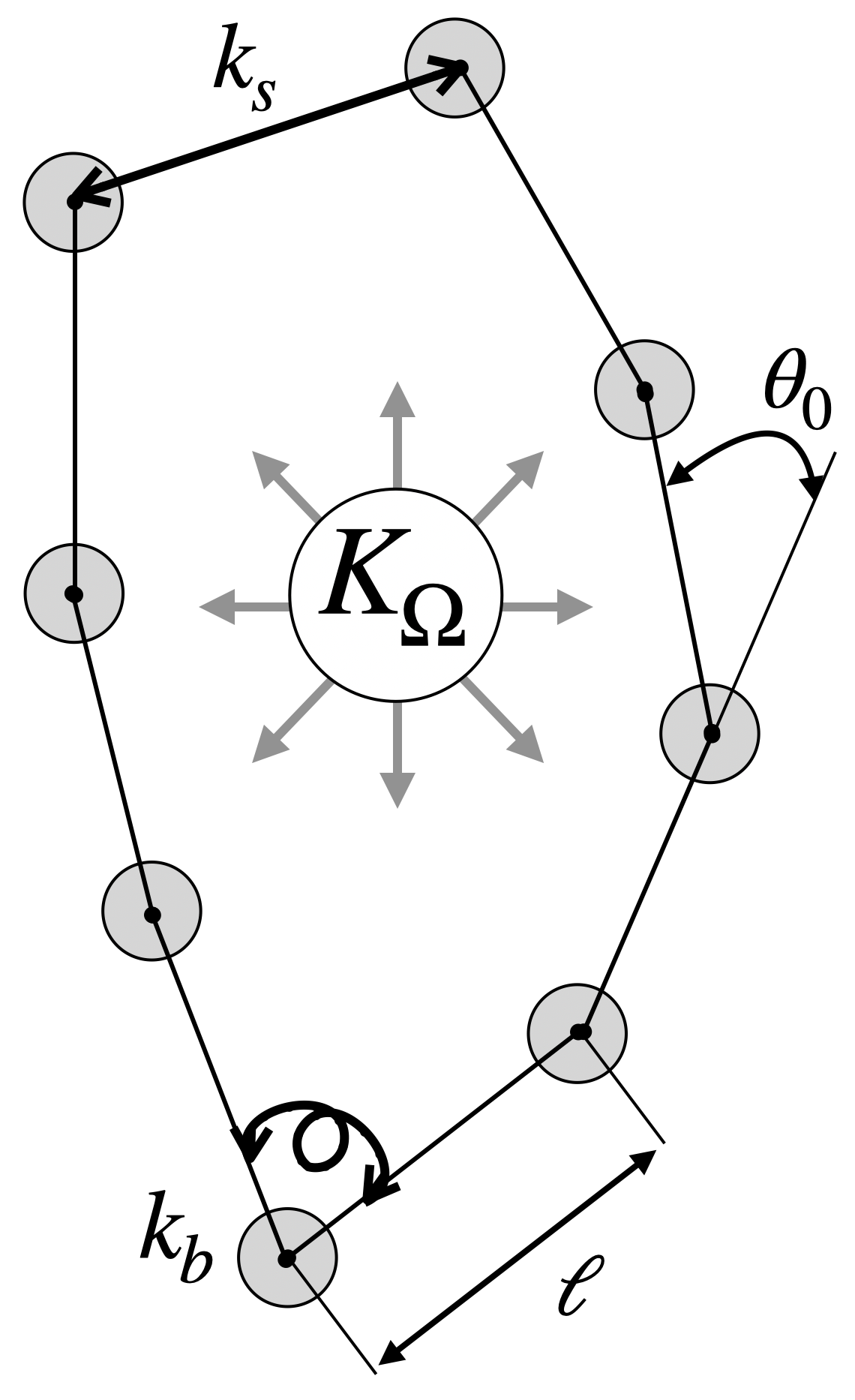}
\caption{Representation of a particle in the $xy$ plane composed of 6 mass points and 
6 segments.}
\label{fig:internal}
\end{figure}

\subsection{Geometric representation of particles} 

In the model, we refer to various basic units—such as plant cells, grains, 
microcapsules, nanotubes, and vesicles—as "particles." 
Each particle can undergo significant elastic 
or plastic deformations without breaking and interacts with other particles through 
elastic and cohesive contact forces. These deformations involve changes in both 
the volume and shape of the particles, which are modeled by representing each 
particle's periphery as an array of mass points connected by elastic-plastic force laws. 
These force laws can be viewed as extensible elements (beams) linking the 
mass points; however, the mass and dynamics of the system are carried by the 
mass points themselves, which we will also refer to as nodes or vertices.
It is important not to confuse this approach with alternative methods that 
represent the particle surface as an array of beam elements articulated 
at their endpoints, where the mass and dynamics are carried by the 
elements rather than the nodes.

In our model, the mechanical integrity of a particle is maintained through 
the interactions between adjacent pairs of mass points and by a flexural   
or bending stiffness at each vertex formed by two adjacent segments (elements), 
which involves three mass points. These mass points at the vertices are the 
discrete elements of the model and represent the shape degrees of freedom 
of a particle. Changes in particle shape are governed by variations in the 
angles between adjacent segments (the bending mode), 
while variation of the particle's perimeter depends on the elongation 
of these segments (the stretch mode).  
Although the dynamics is carried by the vertices, we assume that for 
interactions between particles, the segments behave as 
sphero-segments—thick lines with rounded caps—that can come 
into contact with the vertices of other particles.
  
Figure~\ref{fig:internal} illustrates a 2D particle consisting of vertices and elements 
of varying lengths $\ell$ in the $xy$ plane. The initial shape of the particle 
is defined by the equilibrium lengths $\ell_0$ of its sides and the angles 
$\theta_0$ between adjacent elements. The corresponding stretch and 
bending stiffnesses are denoted by $k_s$ and $k_b$, respectively. 
The particles are assumed to have a length $s$ along the $z$ axis.  
The set of vertices defines a thin shell, but we numerically replace each 
vertex by a disk of radius $r$ and each segment joining two vertices by 
a sphero-segment or rounded-cap rectangle of thickness $2r$. 
The centers of the disks at the vertices of two adjacent elements coincide.

\begin{figure}[tbh]
\centering
\includegraphics[width=0.96\columnwidth]{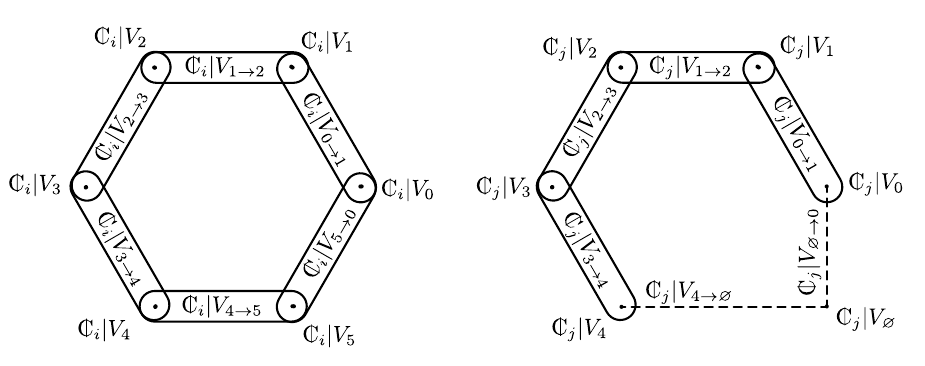}
\caption{
Identification of vertices and segments in each particle. 
Each vertex has a local number within the particle; to refer to it 
globally, both the particle number and the local vertex number are needed. 
For instance, the bottom-right vertex is identified as $\Cell{j}{4}$, where 
$j$ is the particle number and 4 is the local vertex number. Similarly, any vertex 
in particle $i$ is denoted as $\Cell{i}{k}$, where $k$ specifies the vertex within that particle.}
\label{fig:indices}
\end{figure}

We employ a local numbering scheme for vertices within each particle 
to ensure precise identification. As illustrated in Fig. \ref{fig:indices}, 
this method assigns each vertex a pair of indices, denoted as $\Cell{i}{k}$, 
where $i$ represents the particle number and $k$ indicates the local 
vertex number. An element connecting two successive vertices $k$ and $k+1$ 
within the same particle $i$ is labeled as $\Cell{i}{k\rightarrow k+1}$.
Additionally, our data structure accounts for cases where an element may 
have a missing endpoint, which is useful for modeling open or breakable cells, 
though such cases are beyond the scope of this paper. 
To accommodate these scenarios, we introduce the concept of 
an ``absent" vertex, denoted by $\varnothing$, as seen 
in $\Cell{i}{4 \rightarrow \varnothing}$. This linked list structure, whether 
closed or open, is both concise and easy to implement.

\subsection{Internal forces and moments}

Two mass points $k$ and $k+1$ are assumed to interact through 
a harmonic potential energy with a plastic threshold. In other words,  
the extensional incremental deformation  $\delta\ell_{(k,k+1)}$ of 
the segment $\Cell{i}{k\rightarrow k+1}$ joining the two points 
is governed by a linear elastic-perfectly plastic law: 
\begin{equation}
\delta F_{(k,k+1)} = \left \lceil k_s \, \delta\ell_{(k,k+1)} \right \rfloor_{\displaystyle \pm F_s^\text{Y}}, 
\label{eqn:F}
\end{equation}
where $k_s$ is the stretch (extensional) stiffness, $\delta F_{(k,k+1)}$ is the  
extensional force increment, and $F_s^\text{Y}$ is 
the elastic limit or plastic threshold force.   
The notation  $\left \lceil \Delta \mathcal{X} \right \rfloor_{\displaystyle \pm \mathcal{P}}$ 
is defined as:
\begin{equation}
\left \lceil \Delta \mathcal{X} \right \rfloor_{\displaystyle \pm \mathcal{P}} = 
\left\{
\begin{array}{ll}
\Delta \mathcal{X}  &\mbox{if} \ \  \vert \mathcal{X} + \Delta  \mathcal{X} \vert <  \mathcal{P}, \\
0  &\mbox{otherwise.} 
\end{array}
\right.
\label{eqn:X}
\end{equation}
In this notation, compressive forces and contractions of a  segment  
are counted as positive, whereas tensile forces and extensions 
are considered negative.

Shear deformation across the shell wall is neglected. 
However, to endow a particle with an equilibrium shape, we must also 
incorporate resistance to bending at all vertices of the particle. 
We assign a reference angle $\theta_{0k}$ 
to each node $\Cell{i}{k}$,  
and the variations $\delta \theta_k$ of the angles follow a linear elasto-plastic law:  
\begin{equation}
\delta M_k = \left \lceil k_b \, \delta \theta_k \, \right \rfloor_{\displaystyle \pm M_b^\text{Y}}, 
\label{eqn:M}
\end{equation}
where $k_b$ is the bending stiffness, $\delta M_k$ is the  
torque increment at  node $k$ due to internal or  
external forces, and $M_b^\text{Y}$ is the  elastic limit or plastic threshold 
for the torque.    
The initial values $\ell_{0(k,k+1)}$ of segment lengths and  
$\theta_{0k}$ of  vertex angles define the particle's initial shape.  
Any changes in shape require external forces. In the elastic regime, 
the initial shape represents the reference state such that if external forces are 
removed, the particle will return to this initial shape.

When plastic deformations occur at some nodes under the action of external forces, the torque 
$M_k$ at those nodes remains equal to  $M_b^\text{Y}$ while the angles  
and consequently the shape of the particle change. In the same way, 
an element may reach its extensional elastic limit, in which case the force will remain at its 
plastic limit $F_s^\text{Y}$ unless the direction of loading changes. Once 
external forces are removed, the new reference state is defined by 
the cumulative plastic deformations at all nodes and elements.  
As we shall see in section \ref{subsec:scheme}, 
since our model assigns degrees of freedom only to mass points, an 
interaction law like  (\ref{eqn:M}), which involves the angle $\delta \theta_k$ as a 
variable, represents a three-body interaction between the three nodes 
$k-1$, $k$, and $k+1$. 
This interaction is handled by substituting the 
torques with moments of forces acting on the segments.     

To accurately model the physical behavior of systems such as 
cellular tissues, it is crucial to account for the effect of fluid inside the particle.
Let $V$ denote the volume of the particle, $V_0$ its initial volume, and $p$ the fluid pressure. 
We assume that the fluid is inviscid, and the variations $\delta p$ of pressure  
follow a linear isotropic elastic behavior: 
\begin{equation}
\delta p = -K_\Omega \dfrac{\delta V}{V_0},   
\label{eqn:Ki}
\end{equation}
where  $K_\Omega$ is the core modulus due to the presence of the fluid. 
While the fluid can possess viscosity or exhibit more complex behaviors, 
this model assumes an inviscid fluid for simplicity.   
Under undrained conditions, where the fluid is confined within the particle, 
the fluid pressure $p$ adjusts in response to changes in the particle's volume. 
Conversely, under drained conditions, where the fluid is in equilibrium 
with the surrounding environment, $p$ remains constant and equal to the external pressure. 
As discussed further below, the overall bulk modulus $K$ of a particle is a combination of 
$K_\Omega$ and $k_s$.

\begin{figure}[tbh]
\centering
\includegraphics[width=0.5\columnwidth]{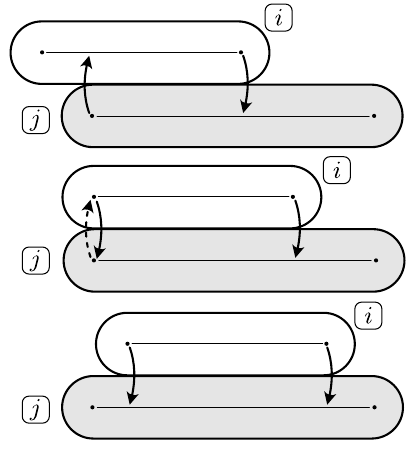}
\caption{An element-element contact can involve either one vertex-vertex contact, 
indicated by two arrows, along with one vertex-element contact, 
or it can involve two vertex-element contacts, each marked by a single arrow.}
\label{fig:ee}
\end{figure}

\subsection{Particle interactions}
\label{subsec:interactions}

In 2D, the interactions between two particles $\Cellonly{i}$ and $\Cellonly{j}$ 
reduce to interactions between their vertices and elements. 
A vertex-vertex contact  occurs between two disks, 
while a vertex-element contact is a contact between a disk and a thick line. 
An element-element contact can involve either one vertex-vertex contact 
along with one vertex-element contact, or two vertex-element contacts, 
as illustrated in Fig. \ref{fig:ee}. Therefore, for detecting the contacts between 
two particles, it is sufficient to consider only vertex-vertex and 
vertex-element interactions. 

\begin{figure}[htb]
\centering
\includegraphics[width=0.5\columnwidth]{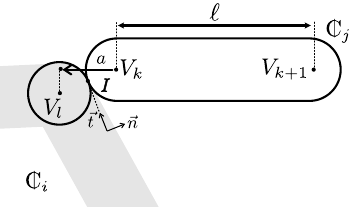}
\includegraphics[width=0.5\columnwidth]{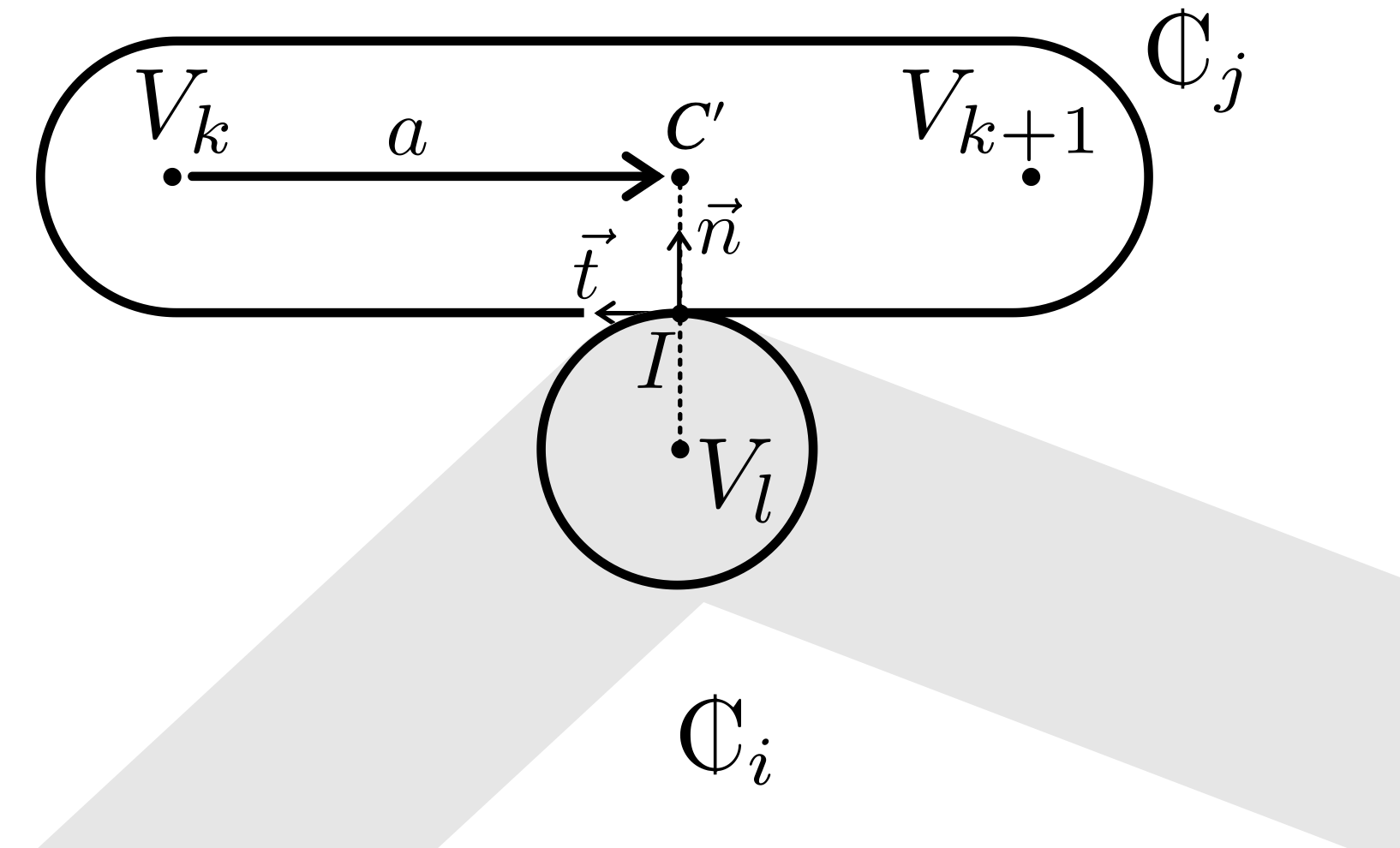}
\includegraphics[width=0.5\columnwidth]{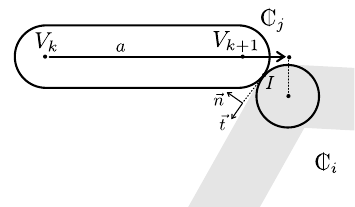}
\caption{Possible contact configurations between two elements.}
\label{fig:3cases}
\end{figure}

Figure \ref{fig:3cases} illustrates the three possible interactions 
between a vertex of $\Cellonly{i}$ and $\Cellonly{j}$: 
$\Cell{j}{k}$-$\Cell{i}{k'}$ (vertex-vertex), 
$\Cell{j}{k}$-$\Cell{i}{k'\rightarrow k'+1}$ (vertex-element), 
and $\Cell{j}{k}$-$\Cell{i}{k'+1}$ (vertex-vertex). 
These cases can be algebraically distinguished by analyzing the 
length $a = \overrightarrow{AC}\cdot\overrightarrow{AB}/\|\overrightarrow{AB}\|$ 
and considering the different in-plane thicknesses 
$2r_i$ and $2r_j$ of cells $\Cellonly{i}$ and $\Cellonly{j}$:
\begin{enumerate}
    \item for $a<0$, the normal vector at contact is 
    $\vec n = \overrightarrow{CA}/\|\overrightarrow{CA}\|$ and the overlap is 
    $\delta_n = \|\overrightarrow{CA}\|-\left(r_i+r_j\right)$, 
    \item for $0\leqslant a \leqslant \ell$, $\vec n = \overrightarrow{CC'}/\|\overrightarrow{CC'}\|$ and $\delta_n = \|\overrightarrow{CC'}\|-\left(r_i+r_j\right)$, 
    \item for $a > \ell$, $\vec n = \overrightarrow{CB}/\|\overrightarrow{CB}\|$ and $\delta_n = \|\overrightarrow{CB}\|-\left(r_i+r_j\right)$.  
\end{enumerate}
Here, for a contact $I$ between the vertex $\Cell{j}{k}$ and 
the element $\Cell{i}{k'\rightarrow k'+1}$, $C$ is the center,  
and $C'$ is the projection of $C$ over the line $(AB)$ 
defined by the two endpoints $A$ and $B$ of the element. 
We also denote by $\left(I,\vec n,\vec t\right)$ the local frame attached to the contact point 
with $\vec t = \vec n \times \vec z$. 

The total contact force $\vec f$ at the contact $I$ is defined by its normal 
and tangential components $f_n$ and $f_t$ in the local frame: 
\begin{equation}
\vec f = f_n \vec n + f_t \vec t.
\end{equation}
In all cases, the normal elastic repulsion force is given by:
\begin{equation}
f_n = k_n \delta_n, 
\end{equation}
where $k_n$ is the normal contact stiffness. 
The tangential elastic force follows the Coulomb friction law, 
which can be expressed using force increments $\delta f_t$ as follows: 
\begin{equation}
\delta f_t = \left \lceil -k_t\, \delta u_t\, \right \rfloor_{\displaystyle \pm \mu f_n}, 
\end{equation}
where $\mu$ is the friction coefficient, $\delta t$ is the time step, $k_t$ is the 
tangential elastic stiffness, and $\delta u_t$ is the incremental tangential 
displacement of $\Cellonly{j}$ with respect to 
$\Cellonly{i}$ at the contact point.  

\subsection{Time stepping scheme}
\label{subsec:scheme}

Given that the discrete degrees of freedom for the particles are carried 
by their vertices, we assign a mass $m$ to each vertex, resulting in a 
total particle mass of $Nm$ for a particle composed of $N$ vertices. 
The classical velocity Verlet time-stepping scheme is applied to each mass point. 
This scheme, widely used in DEM, is recognized for its balance 
between accuracy and computational cost. As a symplectic integrator, 
it conserves the phase-space volume over time, making it well suited 
for Hamiltonian systems. Unlike DEM, rotational degrees of freedom for 
the mass points are not required, as particle rotations naturally arise from 
the displacement of vertices.
The time step $\delta t$ must be kept below the critical 
threshold $\delta t_c = \sqrt{m/k_{\max}}$, where $k_{\max}$ 
represents the largest stiffness parameter of the system. 

\begin{figure}[tbh]
\centering
\includegraphics[width=0.6\columnwidth]{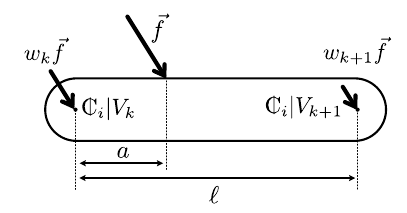}
\caption{Representation of force transfer from the contact point to the nodes belonging 
to $\Cell{i}{k \rightarrow k+1}$.}
\label{fig:force_transfert}
\end{figure}

Although the elements are massless, they can still receive  
forces at their contact points with other particles. Since the dynamics is  
entirely governed by the mass points located at the vertices, 
any force acting on an element must be transferred to the vertices 
at its endpoints, as illustrated in Fig.~\ref{fig:force_transfert}. 
Let $\vec{f}$ represent a force vector acting on an element of length $\ell$, 
applied at a point located a distance $a$ from one end, denoted as $k$, 
and a distance $\ell - a$ from the other end, denoted as $k+1$. 
Assuming the element is in equilibrium under the action of $\vec{f}$ and 
the reaction forces at the two vertices, the force $\vec{f}$ is fully 
transferred to these endpoints. The forces acting at the vertices $k$ and $k+1$ are given 
by $\vec{f}_k = w_k \vec{f}$ and $\vec{f}_{k+1} = w_{k+1} \vec{f}$, 
respectively, where $w_k = (\ell - a)/\ell$ and $w_{k+1} = a/\ell$.

The core pressure within a particle manifests through the force it 
exerts on each vertex. This process occurs in two distinct steps. 
First, a force increment $\delta \vec{F}^{\text{in}}_{(k,k+1)}$ due to the 
core pressure increment $\delta p$ acts on each element $(k,k+1)$. 
Given the isotropic nature of the pressure, this force increment is perpendicular 
to the element at its midpoint and is calculated as $s \ell_{(k,k+1)} \delta p$, 
where $\ell_{(k,k+1)}$ is the length of the element.
Second, this force is equally distributed between the two vertices $k$ and $k+1$, 
in accordance with the rule outlined in Section \ref{subsec:interactions}. 
The variation $\delta p$ in core pressure during each time step is 
given by the corresponding volume change, which is based 
on the displacements of the vertices at the midpoint of each time step, 
similar to how contact forces with other particles are handled.

\begin{figure}[tbh]
\centering
\includegraphics[width=0.98\columnwidth]{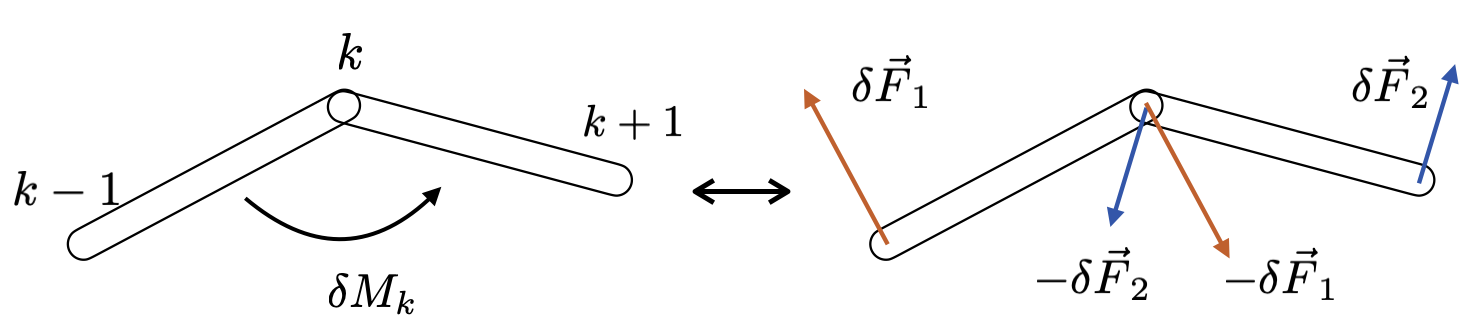}
\caption{Equivalence between a torque acting at the node $k$ and 
a pair of couples acting on the elements $(k-1,k)$ and $(k,k+1)$.}
\label{fig:couple}
\end{figure}

As previously mentioned, the torque equation $\eqref{eqn:M}$ 
represents a three-body interaction, given that the primary variables of the system 
are the displacements of the mass points. 
The variation $\delta \theta_k$ of the angle at a vertex $k$ is computed 
at the midpoint of each time step. Using Eq. $\eqref{eqn:M}$, 
the increment $\delta M_k$ is calculated. 
As shown in Fig. \ref{fig:couple}, this torque increment can be replaced by two 
couples $(\delta \vec{F}_1, -\delta \vec{F}_1)$ and $(\delta \vec{F}_2, -\delta \vec{F}_2)$, 
with the force magnitudes given by:
\begin{eqnarray}
\delta F_1 &=& \frac{\delta M_k}{2\ell_{(k-1,k)}} \\
\delta F_2 &=& \frac{\delta M_k}{2\ell_{(k,k+1)}}. 
\end{eqnarray}
The forces $\delta \vec{F}_1$ and $-\delta \vec{F}_1$ are perpendicular to 
the element $(k-1,k)$ and act on the vertices $k-1$ and $k$, respectively. 
Similarly, the forces $\delta \vec{F}_2$ and $-\delta \vec{F}_2$ are perpendicular 
to the element $(k,k+1)$ and act on the vertices $k$ and $k+1$, respectively. 
The sum of the two couples equals $\delta M_k$.

Finally, a combination of forces and velocities can be imposed on the particles 
through their vertices. Specifically, a particle can be made perfectly rigid by 
freezing its surface degrees of freedom. In the same way, wall boundary conditions 
can be enforced  by imposing force or velocity components 
on selected vertices of a particle.

\section{Calibration and verification}
\label{sec:verif}

The parameters of the force laws used in the model must be expressed 
as functions of the material properties and the number $ N $ 
of segments or their lengths $\ell$. We  assume that each particle represents 
a core-shell tube of shell thickness $h$ and length $s$. If  
the shell is made of a material of Young modulus $E$ and Poisson ratio $\nu$,   
according to shell theory,  
the extensional stiffness is given by \cite{fery2007mechanical}
\begin{equation}
k_s = \frac{E h}{1-\nu^2}, 
\label{eqn:k_s}
\end{equation}   
whereas the bending rigidity is 
\begin{equation}
k_b = \frac{E h^3}{12(1-\nu^2)}.
\label{eqn:k_b}
\end{equation}
This expression is based on the consideration that the 
moment $ M_b $ at a vertex is equivalent to the 
bending moment of a continuous beam at the location of the vertex. 
Note that the bending stress over the cross section is given 
by $ \sigma_b = \frac{M_b y}{I_c} $, 
where $ I_c = h^3 \ell/12$ is the centroidal moment of inertia of the beam's cross 
section and $ y $ is the distance from the beam's axis \cite{Beer2015}. 

For the plastic thresholds $ F_s^\text{Y} $ and $ M_b^\text{Y} $, 
we assume that they are related to the material tensile plastic 
stress $ \sigma^\text{Y} $ through the following relations:
\begin{equation}
F_s^\text{Y} = hs\sigma^\text{Y},
\end{equation}
and
\begin{equation}
M_b^\text{Y} = \frac{4I_c}{h}\sigma^\text{Y}.
\end{equation}
The latter assumes that the angular plastic threshold is reached 
when the largest shear stress $\sigma_s = \frac{M_b h}{4I_c} $ in 
the section of the beam is equal to the plastic stress threshold $ \sigma^\text{Y} $.

To verify the code and the calibration of the values of $ k_s $ and $ k_b $, 
we simulated the elastic behavior of a cantilever beam (CB) under 
the action of a force $ F $ applied at its free end and a simply 
supported beam (SSB) under the action of a force $ F $ applied 
at its center for an increasing number of nodes. 
The analytical expression for the normalized elastic deflection 
in the CB case is \cite{Beer2015}:
\begin{equation}
\frac{\delta^{CB}(x)}{\delta^{CB}_{max}} = \frac{1}{2} \left( \frac{x}{L} \right)^2 \left(3 - \frac{x}{L} \right),
\label{eqn:CB}
\end{equation}
where $ L $ is the length of the cantilever, and the origin of 
the $ x $-coordinate is the left fixed end of the beam. 
The maximum elastic deflection occurs at the free end of the cantilever and is given by:
\begin{equation}
\delta^{CB}_{max} = \frac{F}{3} \frac{L^3}{E I_c}.
\end{equation}
The analytical expression for elastic deflection in the case of SSB is:
\begin{equation}
\frac{\delta^{SSB}(x)}{\delta^{SSB}_{max}} = 
\frac{16}{5} \frac{x}{L} \left\{ 1 - 2 \left(\frac{x}{L}\right)^2 + \left(\frac{x}{L}\right)^3 \right\},
\label{eqn:SSB}
\end{equation}
with the maximum deflection at the center of the beam given by:
\begin{equation}
\delta^{SSB}_{max} = \frac{5F}{768} \frac{L^3}{E I_c}.
\end{equation}

\begin{figure}[tbh]
    \centering
    \includegraphics[width=0.45\columnwidth]{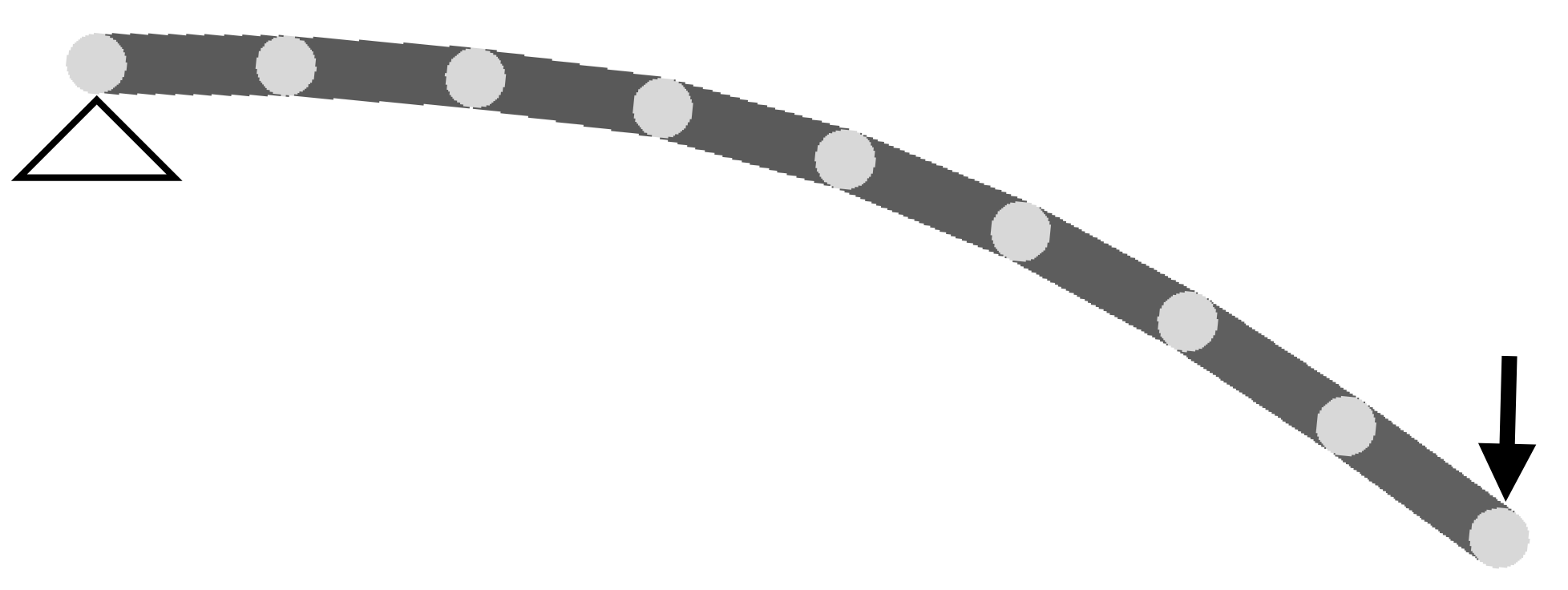}(a)
    \includegraphics[width=0.45\columnwidth]{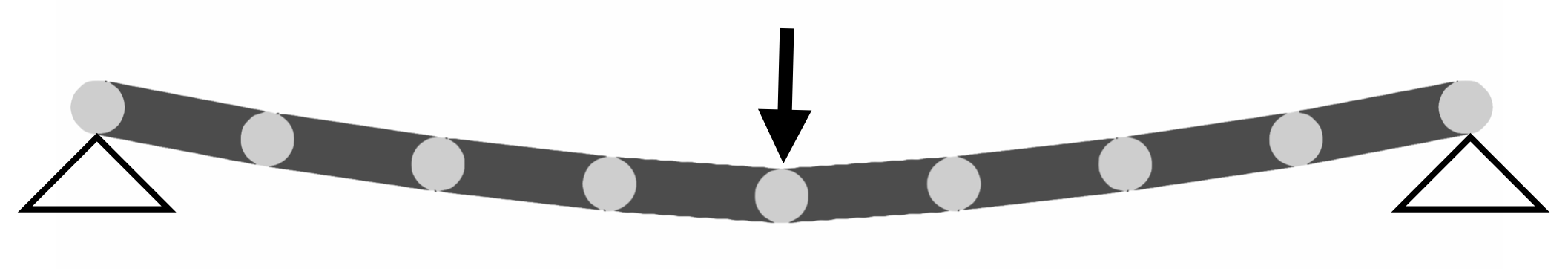}(b)
     
    \includegraphics[width=0.75\columnwidth]{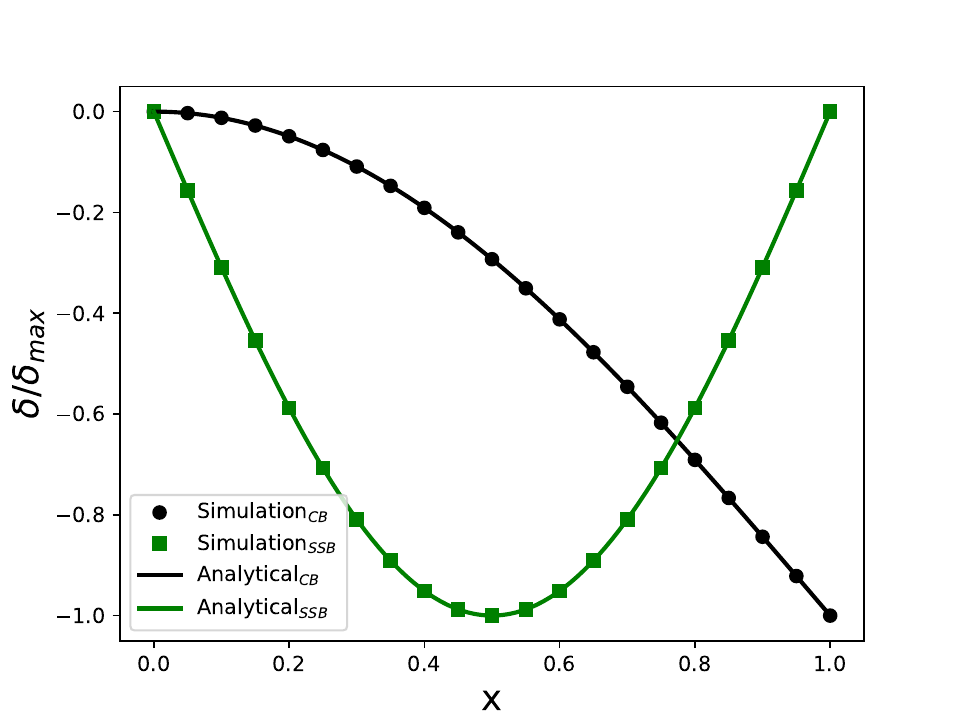}(c)
    \caption{(a) Simulated cantilever beam with 8 elements, (b) simulated 
    simple support beam with 8 elements, and (c) normalized vertical 
    deflection $\delta/\delta_{max}$ as a function of the location $x$ 
    along the beam for cantilever beam (CB) and simple 
    support beam (SSB) with with 20 elements. The solid lines are the 
    analytical expressions (\ref{eqn:CB}) and (\ref{eqn:SSB}). The arrows indicate the location of 
    the external force.}
    \label{fig:CantileverTest}
\end{figure}

Figures \ref{fig:CantileverTest}(a) and (b) display the equilibrium shapes 
of the two beams simulated with 9 nodes, using the expressions (\ref{eqn:k_s}) 
and (\ref{eqn:k_b}) to set the values of $ k_s $ and $ k_b $. 
The vertical deflections $ \delta $, normalized by the maximum deflection $ \delta_{max} $, 
are shown in Fig. \ref{fig:CantileverTest}(b) with 21 nodes, 
alongside the analytical solutions (\ref{eqn:CB}) and (\ref{eqn:SSB}). 
Despite the low number of nodes, the normalized vertical deflections 
are reproduced with high precision for both CB and SSB.

\begin{figure}[tbh]
    \centering
    \includegraphics[width=0.99\columnwidth]{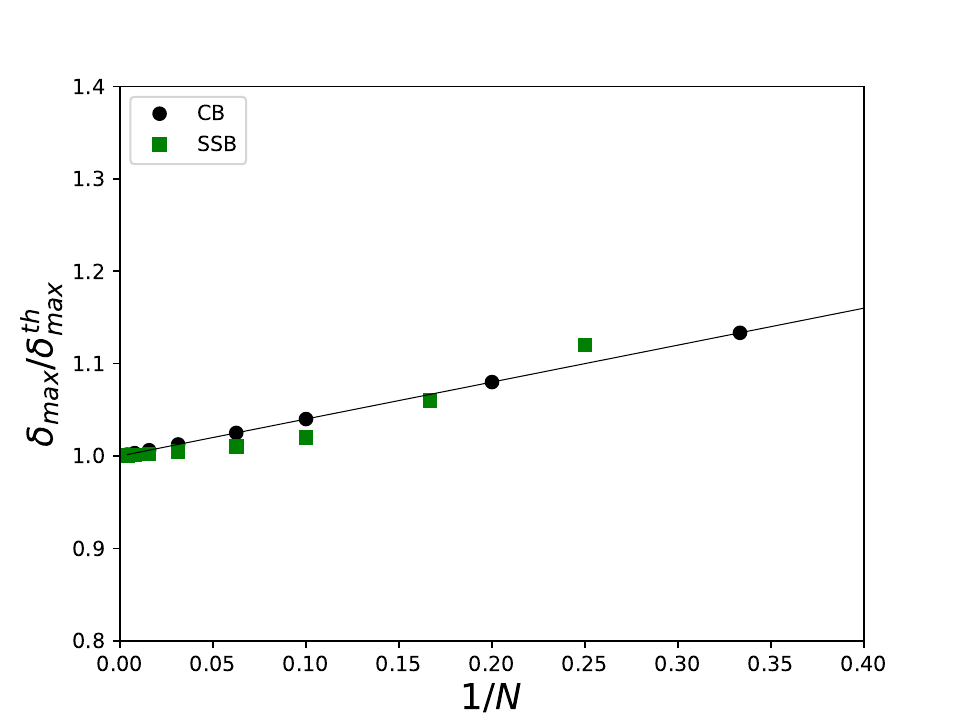}
    \caption{The ratio of the maximum deflection $\delta_{max}$ of 
    cantilever beam and simple support beam to its theoretical value $\delta^{th}_{max}$ as 
    a function of the inverse of the number of nodes $N$.}
    \label{fig:CantileverTestConvergence}
\end{figure}

Figure \ref{fig:CantileverTestConvergence} shows the ratio of $ \delta_{max} $ 
to its theoretical values $ \delta^{th}_{max} $ (standing respectively 
for $ \delta^{CB}_{max} $ and $ \delta^{SSB}_{max} $) as a function of $ 1/N $. 
This ratio is slightly above 1, indicating that the maximum deflection 
is over-estimated by simulations. 
However, as the number of nodes increases and $ 1/N $ tends to zero, 
the ratio converges to 1 as a nearly linear function of $ 1/N $. Hence, 
the numerical solution approaches its analytical value as the number of nodes increases, 
and the value in the continuum limit can be predicted by extrapolation. 
This behavior is a consequence of finite resolution in simulations, 
which affects the strain field, and is therefore independent of the method 
used \cite{Affes2012}.

\section{Diametral compression of a single particle}
\label{sec:compression}

\subsection{Particle-scale parameters}

In our model, a particle is represented as a shell composed of 
mass points with effective properties that depend on the parameters describing the 
interactions between the vertices and the core stiffness. 
Here, we focus on the effective bulk modulus $ K_e $ of the particle, and how the latter 
affects the particle's shape when subjected to diametral compression 
between two flat walls.

\begin{figure}
\centering
\includegraphics[width=0.5\columnwidth]{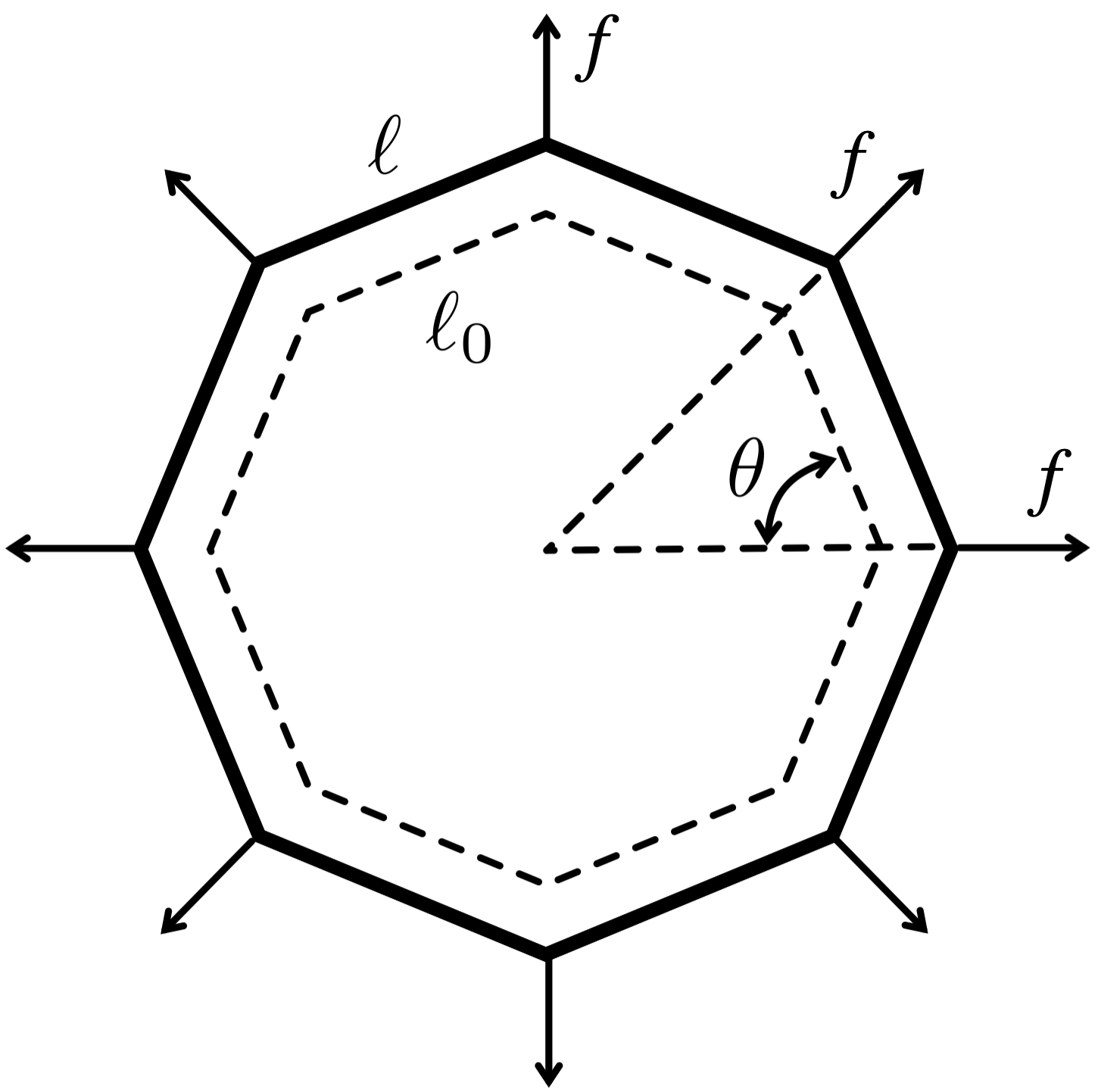}
\caption{A particle composed of $N$ mass points (nodes) and subjected to 
internal pressure $p$.} 
\label{fig:polygone}
\end{figure}

The effective bulk modulus $ K_e $ consists of a contribution $ K_\Omega $ 
from the core and a contribution $ K_{\partial \Omega} $ from the shell:
\begin{equation}
K_e = K_\Omega + K_{\partial \Omega}.
\end{equation}
Let us consider a circular particle of radius $R_0$ 
and subjected to an external pressure $p$, as shown in 
Fig. \ref{fig:polygone}). The application of this pressure leads to the variation 
or the radius from $R_0$ to $R$, the variation $\Delta V = V-V_0$ 
of the volume, and the variation of the length between two mass 
points from $\ell_0$ to $\ell$, such that 
\begin{equation}
    \frac{\Delta V}{V_0} = \frac{2\Delta R}{R_0} = 2 \frac{\Delta \ell}{\ell_0}.
\end{equation}
The pressure is isotropically distributed on the midpoints of the segments. 
Hence, the force acting at each vertex is 
\begin{equation}
f=p \ell s \cos \frac{\pi}{N}.
\end{equation} 
This force is balanced by the extensional force $g = k_s \Delta \ell$ acting 
between the mass points, such that 
\begin{equation}
2g\sin{\frac{\pi}{N}}=f.
\end{equation}
As a result, we have 
\begin{equation}
p = \frac{k_s}{s} \tan \frac{\pi}{N} \frac{\Delta V}{V_0}, 
\end{equation}
implying
\begin{equation}
K_{\partial\Omega} = \frac{k_s}{s} \tan(\frac{\pi}{N}) 
\simeq \frac{2\pi}{Ns} \frac{Eh}{1-\nu^2} 
\label{eqn:domega}
\end{equation}

Using the expression (\ref{eqn:domega}) of $K_{\partial\Omega}$, 
a core-to-shell stiffness ratio can be defined:  
\begin{equation}
\eta = \frac{K_\Omega}{K_{\partial \Omega}}.
\end{equation}
For $ \eta = 0 $, a circular particle behaves as an elastic ring, 
while for $ \eta \gg 1 $, its behavior is expected to be closer to that of a solid particle. 
With Eq. \ref{eqn:domega}, the effective bulk modulus can be expressed as:
\begin{equation}
K_e = (1+\eta) K_{\partial \Omega} \approx (1+\eta) \frac{2\pi}{Ns} \frac{Eh}{1-\nu^2}.
\label{eqn:Ke}
\end{equation}

\begin{figure}[tbh]
    \centering
    \includegraphics[width=0.95\columnwidth]{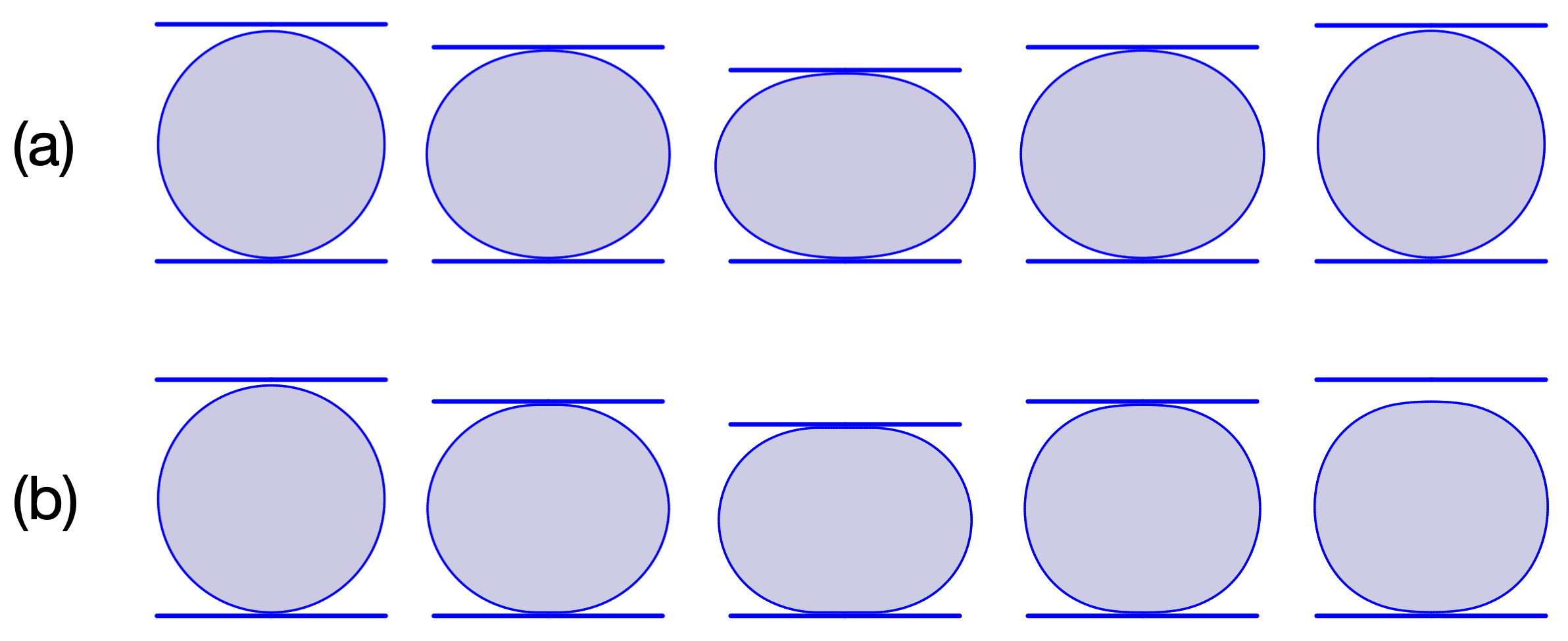}
    \caption{Evolution of particle shape during the compression and decompression  
    of a circular particle between two flat walls for a large value of the stiffness ratio 
    $\eta = \infty$ in the purely elastic case (a) and for a small 
    elastic limit $M_b^\text{Y}$ (b).} 
    \label{fig:CompressCompare}
\end{figure}

We simulated the diametral quasi-static compression-decompression 
cycle of a single circular particle between two flat and parallel walls, 
varying the elastic and plastic parameters of the particle. 
The strain rate is kept sufficiently low to ensure complete relaxation 
after each strain increment. The walls are modeled as rigid bars represented 
by open cells with node reference angles equal to $ 0 $ and with 
high values of $ k_b $ and $ k_s $ compared to those of the particle.
Figure \ref{fig:CompressCompare} illustrates the evolution of the 
particle's shape with a high value of the stiffness ratio $\eta$  
in the purely elastic case (first row)  
and in the case with a small elastic limit (second row), for a 
vertical strain $\varepsilon = 1 - b / b_0 = 0.35 $, 
where $ b $ is the distance between the two walls and 
$ b_0 = 2R $ is its initial value. 
In the purely elastic case, the particle deforms nearly as  
a flattening ellipse, with a minor axis length $ b $ perpendicular to 
the walls and a major axis length $ a $ parallel to the walls. 
Initially, the contacts with the walls are point contacts but develop 
a small flat contact area as deformation continues. 
During unloading, the particle fully recovers its initial circular shape.
In the case with a small elastic limit, the particle rapidly 
reaches the plastic deformation regime during compression, 
resulting in a larger contact area with the walls compared to the purely 
elastic case. We also see that the particle does not fully recover its 
initial shape after unloading as a result of the cumulative plastic deformation 
of the shell.

\begin{figure}[tbh]
    \centering
    \includegraphics[width=0.95\columnwidth]{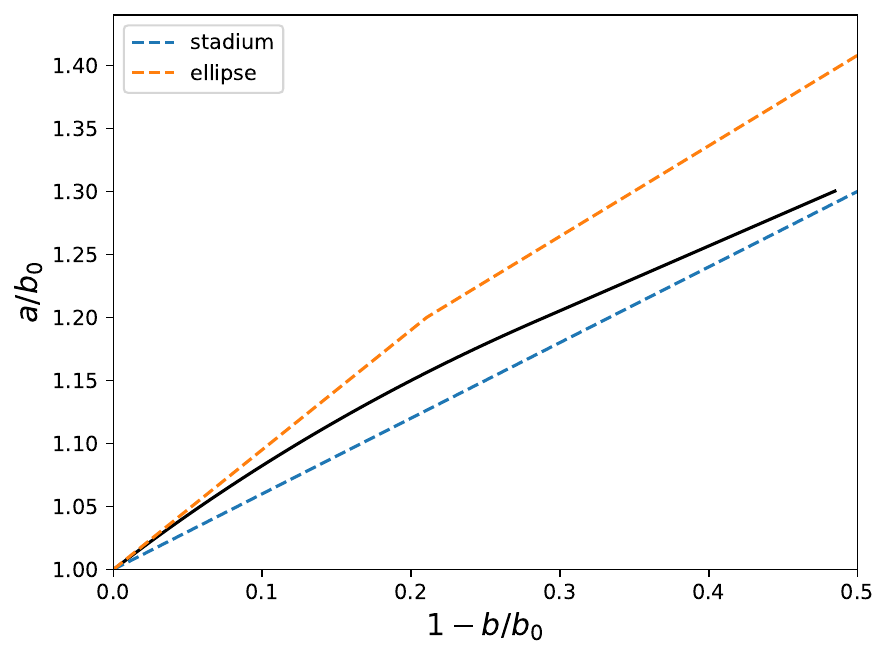}
    \caption{Evolution of the major axis length $a$ normalized by the initial value $b_0$ 
    of the minor axis length as a function of 
    axial strain $1-b/b_0$ during diametral compression of an elastic ring ($\eta=0$) from numerical 
    simulation (black line) and from two models assuming either elliptic (red line) 
    or stadium-like (blue) particle shape with constant perimeter.}
    \label{fig:shapeEvolution}
\end{figure}

\begin{figure}[tbh]
    \centering
    \includegraphics[width=0.95\columnwidth]{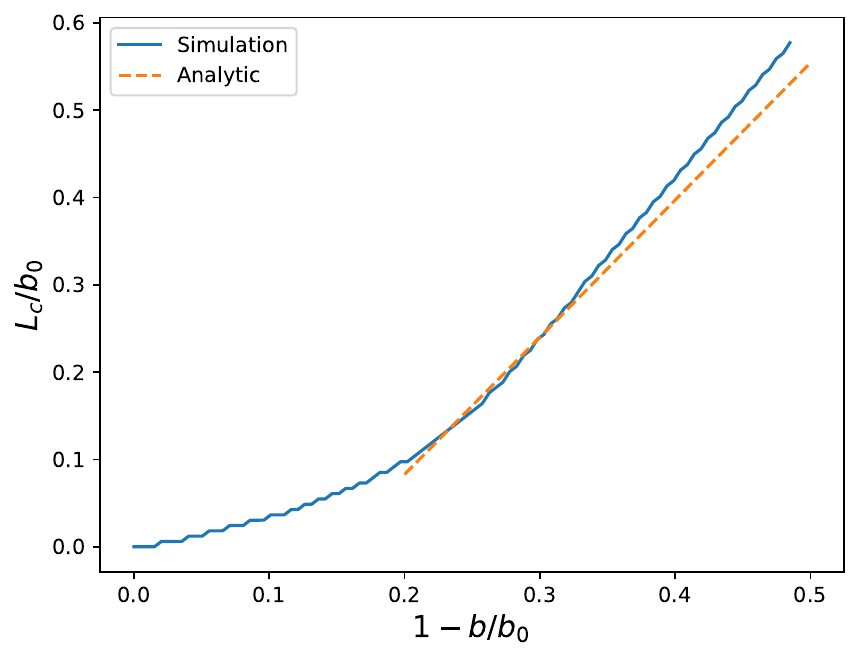}
    \caption{Evolution of the contact length $L_c$ normalized by the initial diameter 
    $b_0$ as a function of axial strain during diametral compression of an 
    elastic ring ($\eta=0$). The slope of the dashed line is $\pi/2$.} 
    \label{fig:ContactLength}
\end{figure}

\subsection{Elastic ring}

At the beginning of compression, the particle flattens but remains rounded. 
As the strain increases, the shape evolves to resemble 
a rounded-cap rectangle or a stadium-like shape, 
characterized by two parallel flat sides in contact with the walls and 
two lateral semi-circular sides \cite{VanHirtum2015}. The evolution of the particle's aspect 
ratio $ a/b $ as a function of $ b $ can be predicted for these two 
shapes under the assumption that the perimeter remains constant. 
For the elliptical shape, the relationship between $ a $ and $ b $ is 
given by the following implicit equation \cite{VanHirtum2015}:
\begin{equation}
2b_0 = 3(a+b) - \sqrt{(a+3b)(3a+b)}, 
\label{eqn:el}
\end{equation}
where $ b_0 $ is the initial distance between the two walls. 
In the case of the stadium-like shape, the aspect ratio $ a/b $ is expressed as:
\begin{equation} 
\frac{a}{b_0} = 1+\left(\frac{\pi}{2} -1 \right) \left(1-\frac{b}{b_0} \right).
\label{eqn:st}
\end{equation} 
These equations provide a simple way to describe the deformation 
behavior of the particle as it is compressed.

\begin{figure}[tbh]
    \centering
    \includegraphics[width=0.95\columnwidth]{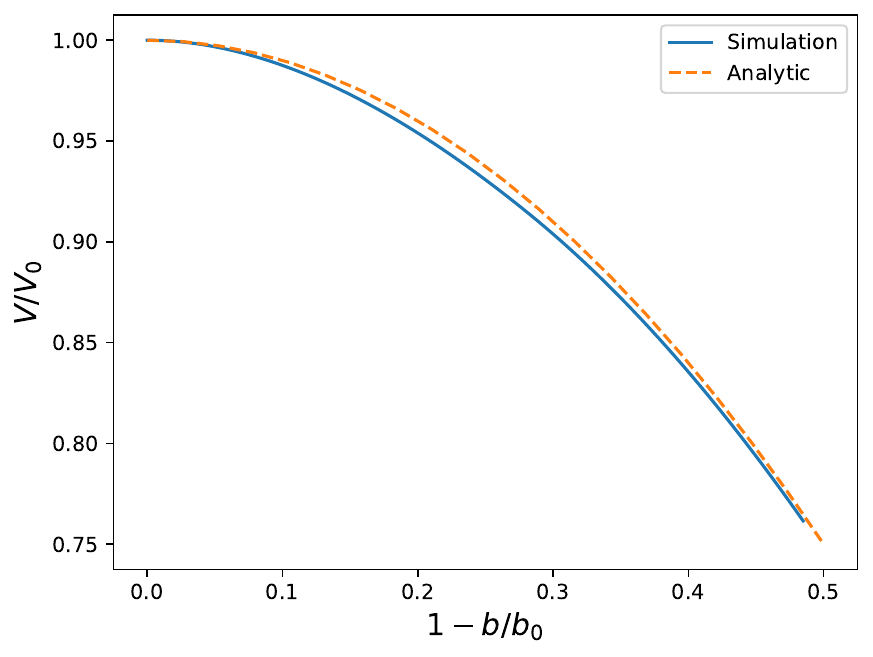}
    \caption{Evolution of the normalized volume of an elastic ring ($\eta=0$) 
    as a function of axial strain during diametral compression.} 
    \label{fig:V}
\end{figure}

The analytical ratio $ a/b_0 $ is plotted as a function 
of $ \varepsilon = 1-b/b_0 $ in Fig. \ref{fig:shapeEvolution} for the two shapes 
discussed earlier, based on equations (\ref{eqn:el}) and (\ref{eqn:st}), 
as well as from numerical simulations with $ \eta=0 $ (elastic ring). 
As expected, the simulated evolution of $ a/b_0 $ falls between the two shapes, 
resembling an ellipse at low $ \varepsilon $ values and transitioning 
toward a stadium-like shape at higher values. 
Figure \ref{fig:ContactLength} illustrates the corresponding evolution 
of the normalized contact length $ L_c/b_0 $. This contact length increases 
nonlinearly with vertical strain but shifts to a linear trend beyond $ \varepsilon \simeq 0.3 $.
The linear behavior in the second regime aligns with the shape transition 
from ellipse to stadium-like. For a stadium-like shape with constant perimeter, 
it can be shown that
\begin{equation}
\frac{L_c}{b_0} = \frac{\pi}{2} \left(1-\frac{b}{b_0} \right).
\label{eqn:Lc-b}
\end{equation}  
A straight line with a slope of $ \pi/2 $ is plotted in Fig. \ref{fig:ContactLength}, 
demonstrating that the slope observed in simulations during the second regime 
is very close to this value.
A similar relation can be derived for the evolution of the volume $ V $ of the 
stadium-like shape under the constant perimeter assumption:
\begin{equation}
\frac{V}{V_0} = 1 - \left(1 - \frac{b}{b_0} \right)^2.
\label{eqn:V}
\end{equation} 
As shown in Fig. \ref{fig:V}, this relation is in excellent agreement with 
the volume evolution of an elastic ring from the onset of compression 
through to large strains.

\begin{figure}[tbh]
    \centering
    \includegraphics[width=0.95\columnwidth]{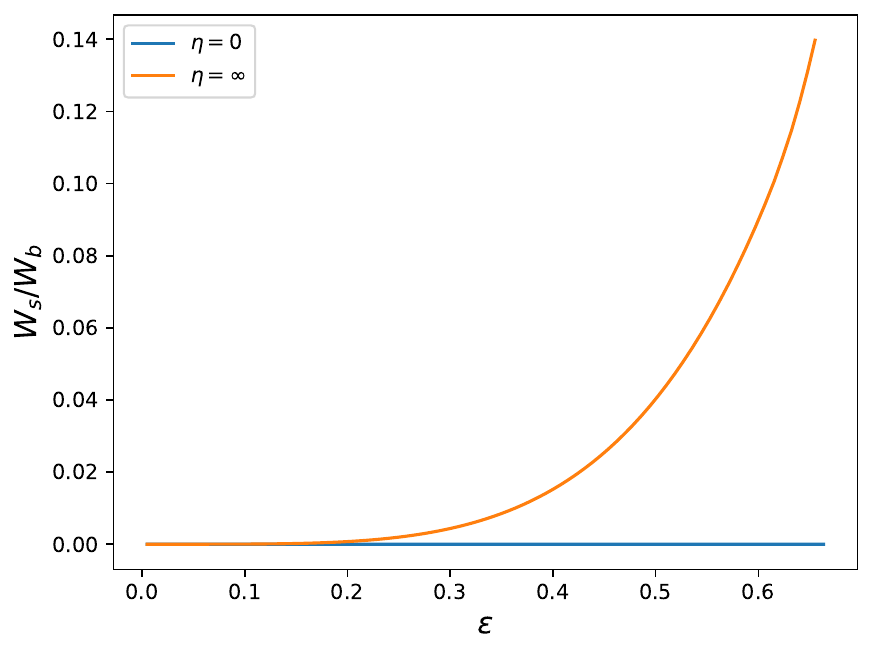}
    \caption{The ratio of extensional energy $W_s$ to bending energy $W_b$ 
    for two values of the stiffness ratio $\eta$ as a function of axial strain.} 
    \label{fig:WsWb}
\end{figure}

\begin{figure}[tbh]
    \centering
    \includegraphics[width=0.95\columnwidth]{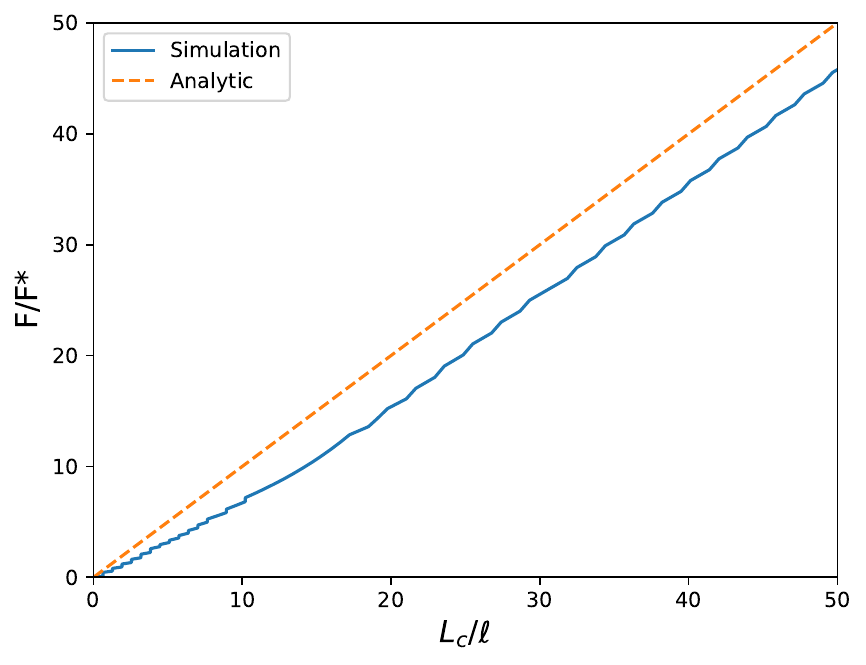}
    \caption{The total wall-particle normal force $F$ normalized by the characteristic force $F^*$ for 
    $\eta=0$ as a function of the contact length $L_c$ during diametral compression of an  
    elastic ring from simulations (full line) and from the approximate analytical 
    expression (\ref{eqn:F-Lc}) (dashed line).} 
    \label{fig:LcF}
\end{figure}

The assumption of a constant perimeter is physically justified 
by the lower energy cost of bending as compared to stretching in the 
case $\eta=0$. In our simulations, we find the ratio of the total stretching energy $W_s$ 
to the bending energy $W_b$ is negligibly small for $\eta=0$ but increases 
with deformation for large values of $\eta$ as shown in Fig. \ref{fig:WsWb}. 
Note that the ratio $W_s/W_b$ 
is proportional to the ratio $k_s/k_b \propto h^{-2}$ and tends to infinity as $h$ 
vanishes.      

It is also interesting to compare the energy costs for the faceting of the particle against the 
walls and for a uniform deformation of the particle. The first deformation mechanism is local 
in the sense that it takes place only at the contact with the two walls while the curvature remains 
equal to the initial curvature everywhere else. The second mechanism involves a slight decrease  
or increase of the curvature at all points of the shell. The total energy stored in 
the two contacts of length $L_c$ with the two walls is $W_{contact} =(k_b/2)(2L_c/R_0^2)$ since 
the curvature changes from $1/R_0$ to 0 and the range of 
deformation is $2L_c$. Alternatively, if the same amount of deformation was   
uniformly distributed on the shell, the change of curvature would be $\pm L_c/R_0^2$ and 
therefore the total energy stored in the shell would be 
$W_{uniform} = (k_b/2) (L_c/R_0^2)^2 (2\pi R_0)$. The latter involves a multiplication 
by the total perimeter $2\pi R_0$ in order to obtain the total energy. Requiring  
$W_{contact} < W_{uniform}$ implies $L_c/b_0 > 1/2\pi \simeq 0.16$. 
This means that the uniform deformation mode is energetically more favorable 
at low strains, but faceting becomes more favorable (with lower energy cost) when the contact 
length exceeds $0.16 b_0$. This is consistent with Fig. \ref{fig:ContactLength}, 
which shows that the contact length increases indeed much faster 
beyond $\varepsilon \simeq 0.25$ corresponding to $L_c \simeq 0.16 b_0$.

Another important variable is the total force $ F $ exerted by the wall on 
the particle. For an elastic ring, the vertical force $ F'_k $ acting on each vertex 
of the particle by the wall creates a local moment $ F'_k \ell $ relative 
to the two neighboring vertices $ k-1 $ and $ k+1 $. 
This moment is balanced by the torque $ M'_k $ resulting from the 
change $ \Delta \theta'_k $ in the vertex angle. 
Initially, the angles are $ 2\pi/N $, but due to compression, the angles 
at the vertices in contact with the wall tend to zero because the 
interface in the contact zone between the wall and the particle is flat. 
Therefore, $ \Delta \theta'_k = 2\pi/N $, leading to $ M'_k = 2\pi k_b/N $. 
The equilibrium of each vertex in the contact zone implies  
$ M'_k = 2\pi k_b/N = F'_k \ell $, which, together with 
Eq. (\ref{eqn:k_b}), leads to $ F'_k = F^* $ with 
\begin{equation}
 F^*= \frac{k_b}{R} = \frac{EI_c}{(1-\nu^2)R\ell}.
 \label{eqn:F*}
\end{equation} 
This force is the characteristic elastic force of the particle. 
Since $ F'_k $ is independent of the vertex position $ k $ in the contact zone, 
the total contact force is the number $ N' $ of nodes on the contact line 
multiplied by $ F^* $. Therefore, for $ N' = L_c/\ell $, we have
\begin{equation}
\frac{F}{F^*} = \frac{L_c}{\ell}.
\label{eqn:F-Lc}
\end{equation}  
The derivation of this expression does not depend on the particle shape, 
suggesting that during diametral compression, Eq. (\ref{eqn:F-Lc}) is valid 
in both small-strain and large-strain regimes.

\begin{figure}[tbh]
    \centering
    \includegraphics[width=0.95\columnwidth]{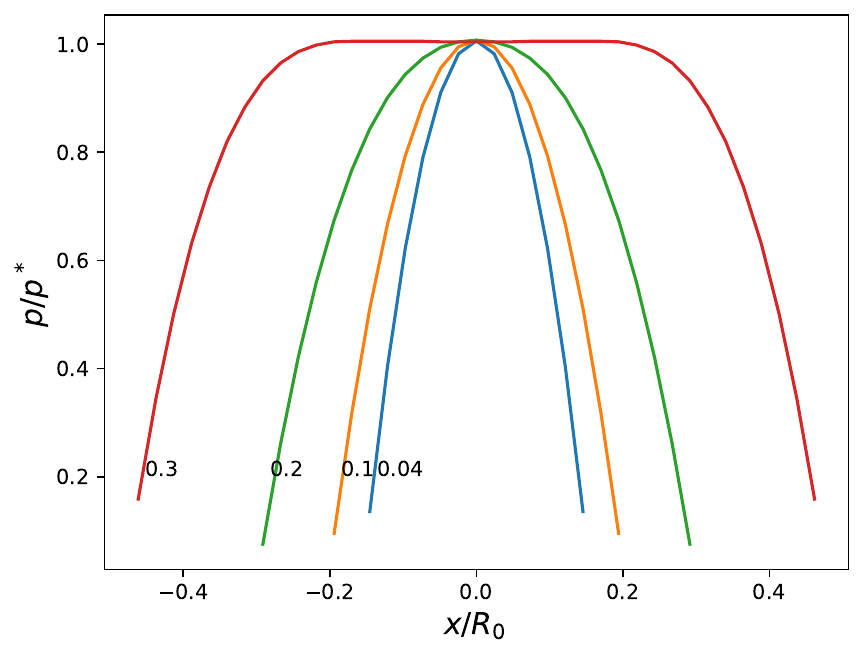}
    \caption{Local pressure $p$ normalized by the characteristic pressure $p^*$ 
    (see text) in the contact zone as a function of nodal position $x$ 
    normalized by the initial particle radius $R$ during diametral compression. 
    The numbers indicate the vertical deformation $1-b/b_0$. The origin of $x$ is the 
    middle of the contact zone.} 
    \label{fig:p-x}
\end{figure}

Figure \ref{fig:LcF} shows $ F/F^* $ for an elastic ring as a function 
of $ L_c/\ell $. The simulation data are close to the expected affine relation 
in both regimes, though with a slightly lower slope in the first regime. 
The simulated forces are consistently below the predicted values 
because the model assumes the contact line is flat and that all forces at 
all nodes in the contact area reach their maximum value $ F^* $. 
An example of the pressure profile along the contact line is shown in 
Fig. \ref{fig:p-x} at four different instants of compression. 
The pressures $ p $ are calculated by dividing the contact force on segments 
of length $ \ell $ by $ s \ell $. The pressures are normalized by the characteristic 
pressure $ p^* = F^*/(s \ell) $. We see that the pressure is not constant across the entire 
contact zone. Initially, the largest pressure $ p^* $ is only reached at the 
center of the contact zone, with a parabolic force profile around it. 
At larger strains, a long plateau of constant pressure $ p^* $ forms, 
but the pressure decreases on both sides of this plateau over several 
nodes, explaining the lower contact force compared to the prediction of Eq. (\ref{eqn:F-Lc}).

\begin{figure}[tbh]
    \centering
    \includegraphics[width=0.95\columnwidth]{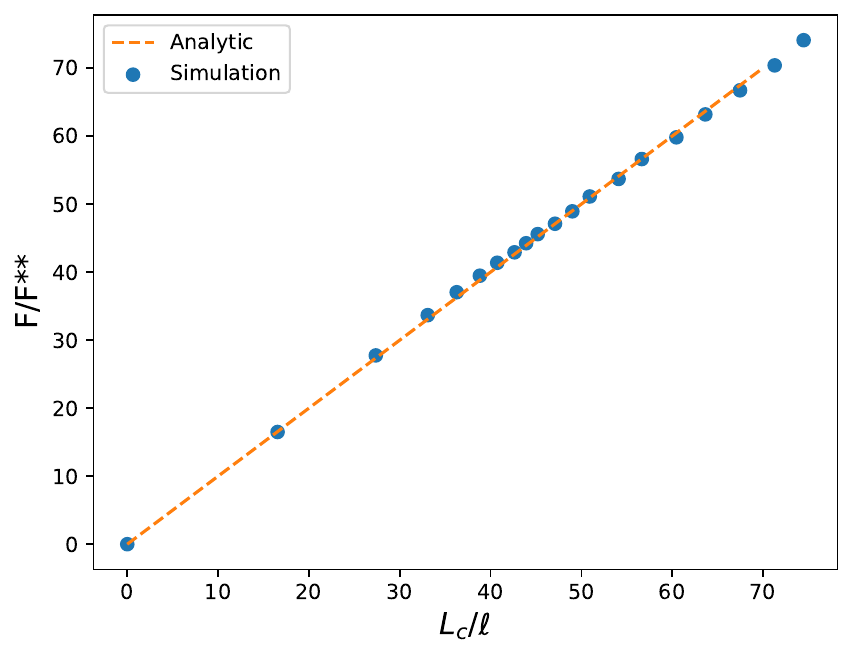}
    \caption{The total wall-particle normal force $F$ normalized by the 
    plastic force threshold $F^{**}$ for 
    $\eta=0$ as a function of the contact length $L_c$ during diametral compression of a   
    plastic ring from simulations (full line) and from the approximate analytical 
    expression (\ref{eqn:F-Lc2}) (dashed line).} 
    \label{fig:F-Lc2}
\end{figure}

\subsection{Plastic ring}

Let us now consider the ring ($\eta = 0$) with a finite plastic 
threshold $M^Y$. This threshold must be compared with the elastic characteristic 
torque $M^*$ defined from Eq. (\ref{eqn:F*}) as follows:
\begin{equation}
M^* = {F^*}{\ell} = \frac{EI_c}{(1-\nu^2)R}.
\end{equation} 
The amount of plastic deformation of a particle depends on the 
ratio $M^Y/M^*$. For illustration, we focus on an example where the plastic 
threshold is reached at very low strains. Simulations of diametral compression 
were conducted for $M^Y/M^* = 5 \times 10^{-5}$. At such a low plastic threshold, 
it can be assumed that all vertices in the contact zone between the particle 
and the driving wall are at the plastic threshold. Following the same reasoning 
as in the case of the elastic ring, all forces are expected to equal the plastic force threshold:
\begin{equation}
F^{**} = \frac{M^Y}{\ell}.
\end{equation}    
Consequently, as in Eq. (\ref{eqn:F-Lc}), the total contact force $F$ is given by
\begin{equation}
\frac{F}{F^{**}} = \frac{L_c}{\ell}.
\label{eqn:F-Lc2}
\end{equation} 

\begin{figure}[tbh]
    \centering
    \includegraphics[width=0.95\columnwidth]{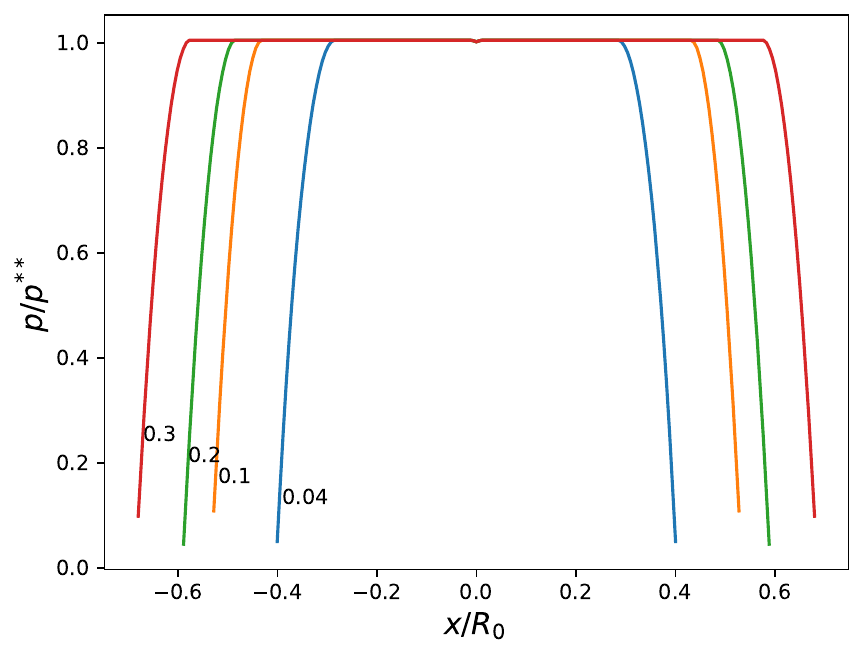}
    \caption{Local pressure $p$ normalized by the characteristic pressure $p^{**}$ 
    (see text) in the contact zone as a function of nodal position $x$ 
    normalized by the initial particle radius $R$ during diametral compression for  
    $M^Y/M^* = 5\times10^{-5}$. 
    The numbers indicate the vertical deformation $1-b/b_0$. The origin of $x$ is the 
    center of the contact zone.} 
    \label{fig:p-x2} 
\end{figure}

Equation (\ref{eqn:F-Lc2}) shows excellent agreement with the simulation 
data, as illustrated in Fig. \ref{fig:F-Lc2}, without distinguishing between 
low- and large-strain regimes. The pressure profile along the contact line is 
displayed in Fig. \ref{fig:p-x2} at four different instants of compression. 
The pressure is normalized by the characteristic pressure $p^{**} = F^{**}/(s \ell)$. 
A broad pressure plateau emerges from the beginning of diametral compression, 
with a steep pressure drop on both sides of this plateau. 
This indicates that the assumption of a flat contact area holds better 
in the plastic case, which explains the excellent agreement of the simulation 
data with Eq. (\ref{eqn:F-Lc2}).

\begin{figure}[tbh]
    \centering
    \includegraphics[width=0.95\columnwidth]{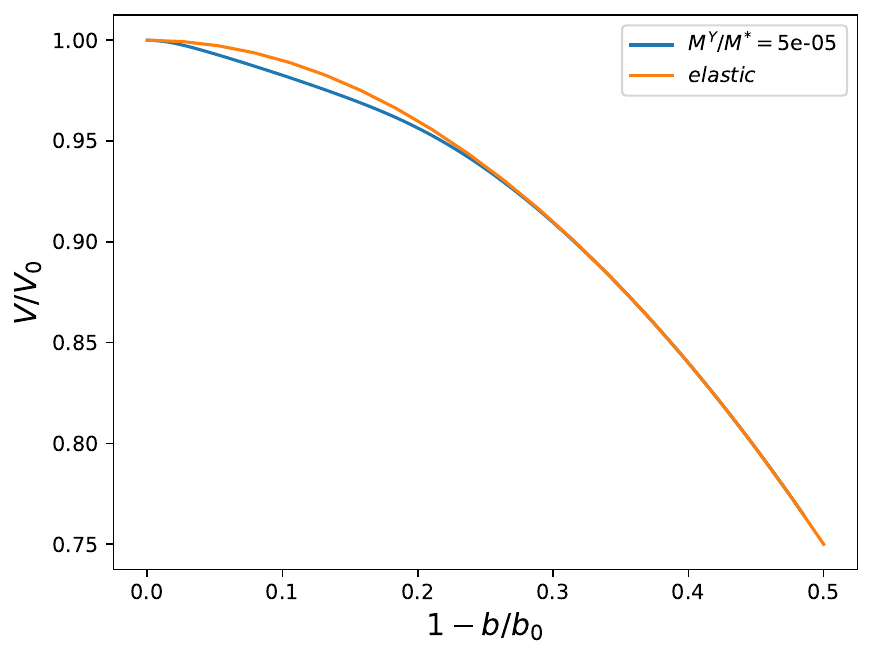}
    \caption{Evolution of the normalized volume of a ring ($\eta=0$)  
    as a function of diametral strain during diametral compression for a plastic ring with 
    $M^Y/M^* = 5\times10^{-5}$ as compared with that of an elastic ring.} 
    \label{fig:V2}
\end{figure}

Figure \ref{fig:V2} displays the particle volume as a function of diametral 
strain $\varepsilon = 1 - b/b_0$. In the plastic case, the volume reduction slightly exceeds 
that of the elastic ring in the initial regime, but they coincide at larger strains. 
The key difference is that the plastic ring retains its distorted shape 
upon unloading, while the elastic ring recovers its original circular shape.

\begin{figure}[tbh]
    \centering
    \includegraphics[width=0.6\columnwidth]{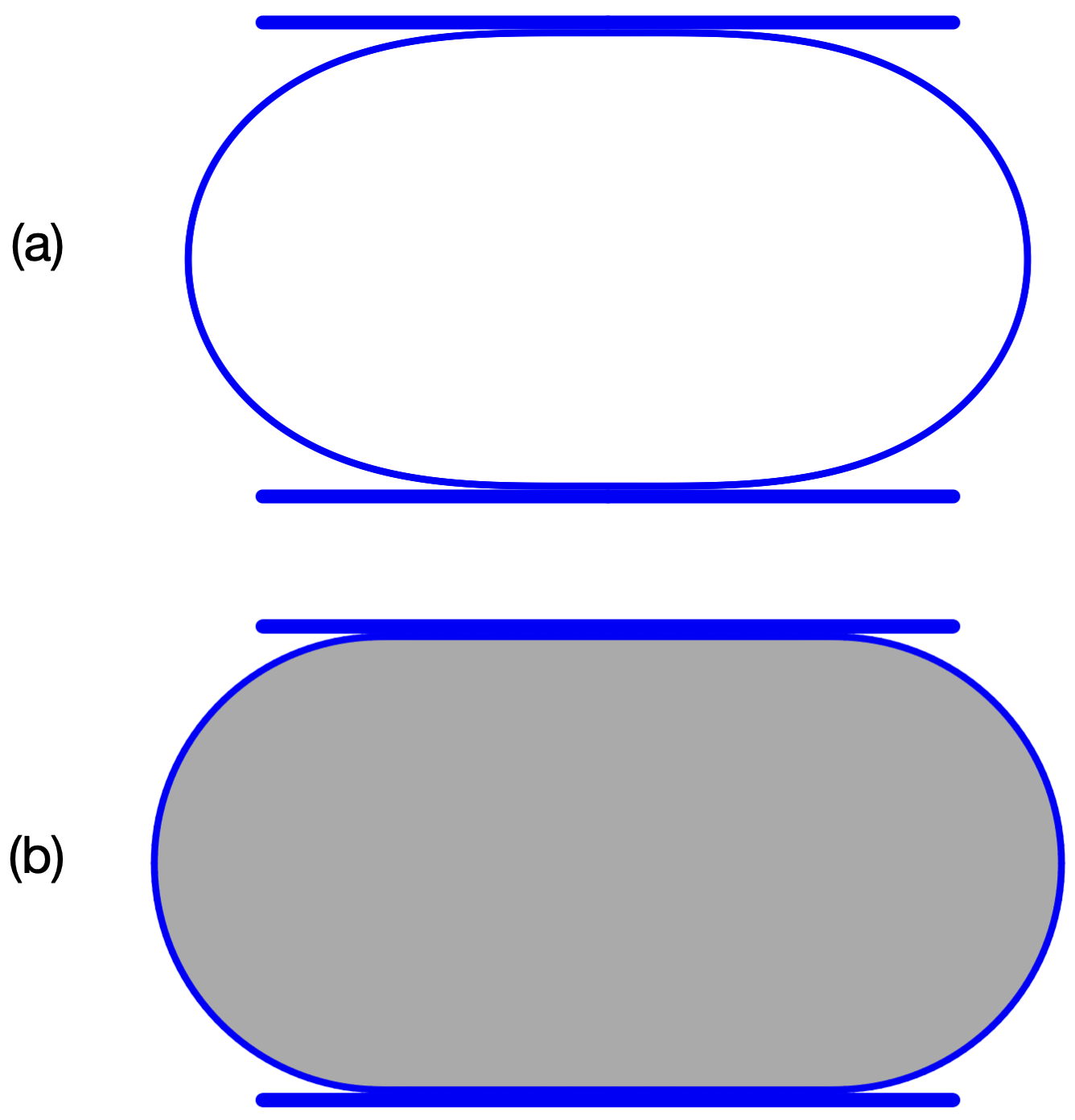}
    \caption{The shape of an elastic ring at diametral strain $1-b/b_0=0.3$ 
    for $\eta=0$ (without core stiffness) (a) and for $\eta=\infty$ (with high core stiffness) (b).} 
    \label{fig:shapes2}
\end{figure}

\begin{figure}[tbh]
    \centering
    \includegraphics[width=0.95\columnwidth]{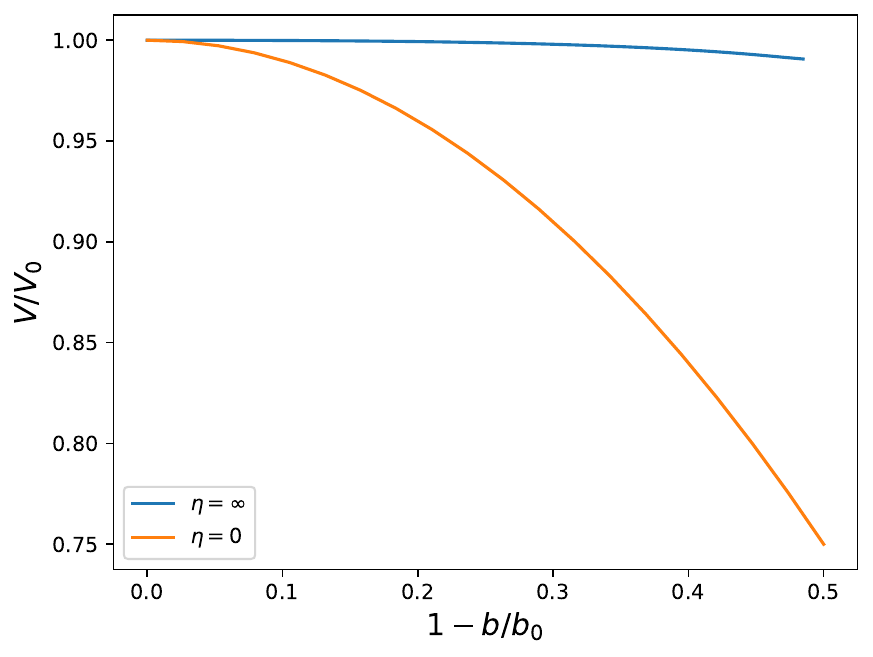}
    \caption{The volume of an elastic ring as a function of diametral strain $1-b/b_0$ 
    for $\eta=0$ (without core stiffness) and for $\eta=\infty$ (with high core stiffness).} 
    \label{fig:V3}
\end{figure}

\subsection{Influence of bulk stiffness}

The presence of a nonzero bulk stiffness $ K_\Omega $ leads to 
an internal pressure $ p $ that arises from volume changes. 
This pressure exerts a radial force $ F_{in} = p s\ell $ on each element, 
which tends to increase the angles between elements, 
especially at the vertices in contact with the wall. We focus here  
on a very large value $ \eta  \simeq 10^5$, denoted as $\eta = \infty$ 
in the figures.

\begin{figure}[tbh]
    \centering
    \includegraphics[width=0.95\columnwidth]{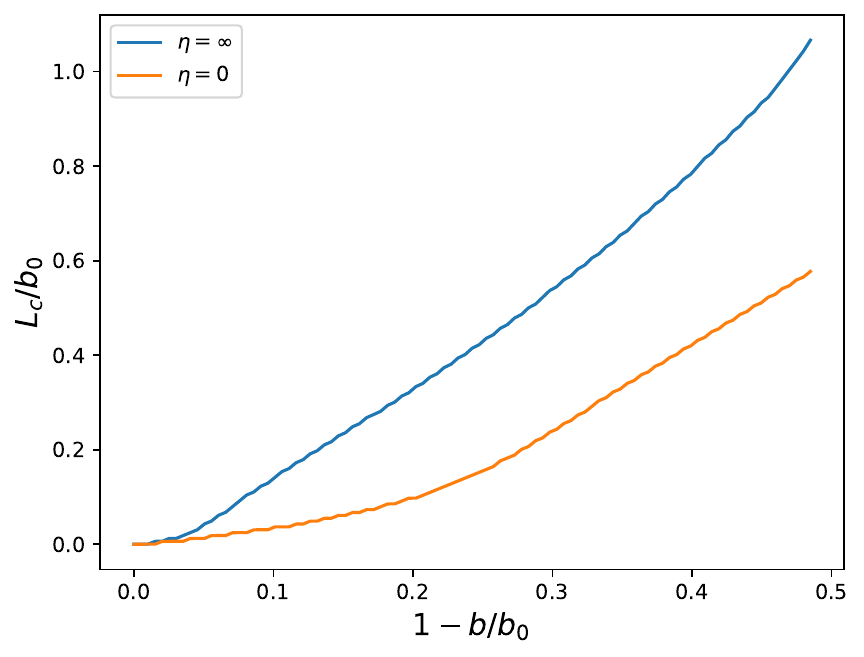}
    \caption{The normalized contact length $L_c$ of an elastic ring as a 
    function of diametral strain $1-b/b_0$ 
    for $\eta=0$ (without core stiffness) and for $\eta=\infty$ (with high core stiffness).} 
    \label{fig:Lc-b3}
\end{figure}

Figure \ref{fig:shapes2} illustrates the shape of an elastic ring at a 
vertical strain of $ 1 - b/b_0 = 0.3 $ for two cases: $ \eta = 0 $ (no bulk stiffness) 
and $ \eta = \infty$.  
As expected, the contact area is significantly larger for $\eta = \infty$. 
In contrast to the $ \eta = 0 $ case, as shown in Fig. \ref{fig:V3}, 
the volume change for $ \eta=\infty$ is minimal and does not follow the 
expression (\ref{eqn:V}) for a stadium-like shape. The typical evolution 
of a stadium-like shape involves a reduction in the radius of the circular sides 
and an increase in contact length. However, the internal pressure 
development hinders the increase in curvature.
As a result, the contact area increases more rapidly, while the curvature 
of the sides increases more slowly compared to the $ \eta = 0 $ case. 
Indeed, Figure \ref{fig:Lc-b3} shows that the growth rate of $ L_c $ is 
approximately 2.1, exceeding the $ \pi/2 $ value expected for a 
stadium-like shape. Consequently, the effect of core pressure or core  
stiffness is to facilitate extensive contacts with the walls (and 
other particles) without a significant change in volume.

\begin{figure}[tbh]
    \centering
    \includegraphics[width=0.95\columnwidth]{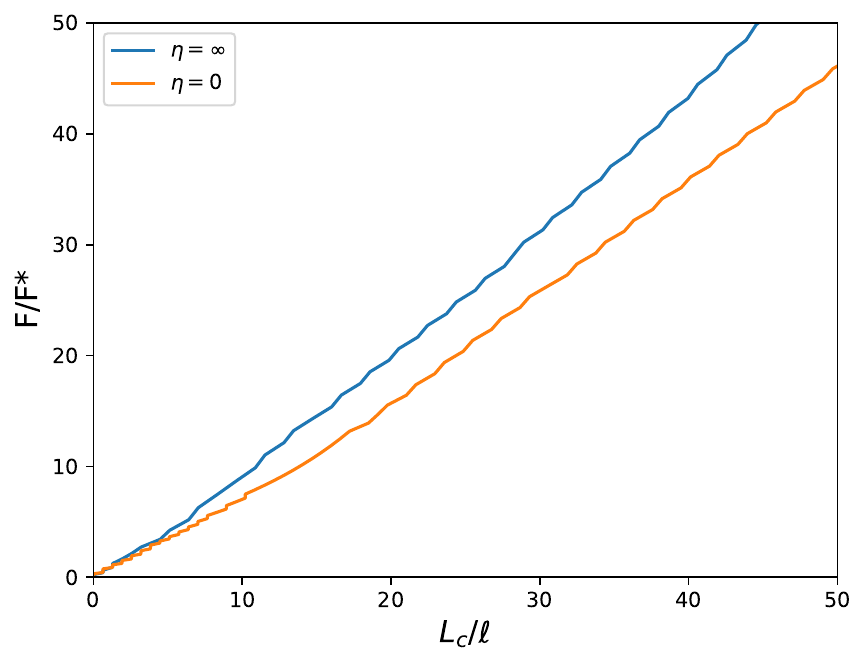}
    \caption{The normalized contact force as a function of contact length $L_c$ 
    during diametral compression of an elastic ring as a function of vertical strain $1-b/b_0$ 
    for $\eta=0$ (without core stiffness) and for $\eta=\infty$ (with high core stiffness).} 
    \label{fig:F-Lc3}
\end{figure}

\begin{figure}[tbh]
    \centering
    \includegraphics[width=0.95\columnwidth]{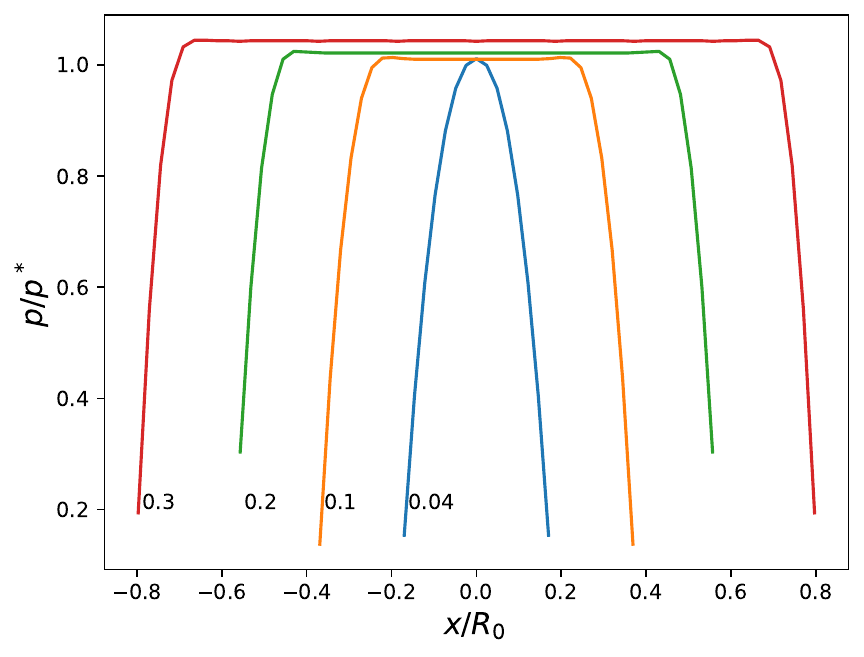}
    \caption{Local pressure $p$ normalized by the characteristic pressure $p^{*}$ 
    (see text) in the contact zone as a function of nodal position $x$ 
    normalized by the initial particle radius $R$ during diametral 
    compression of an elastic ring for  
    $\eta=\infty$ (with high core stiffness). 
    The numbers indicate the diametral deformation $1-b/b_0$. The origin of $x$ is the 
    middle of the contact zone.} 
    \label{fig:p-x3}
\end{figure}

\begin{figure}[tbh]
\centering
\includegraphics[width=0.95\columnwidth]{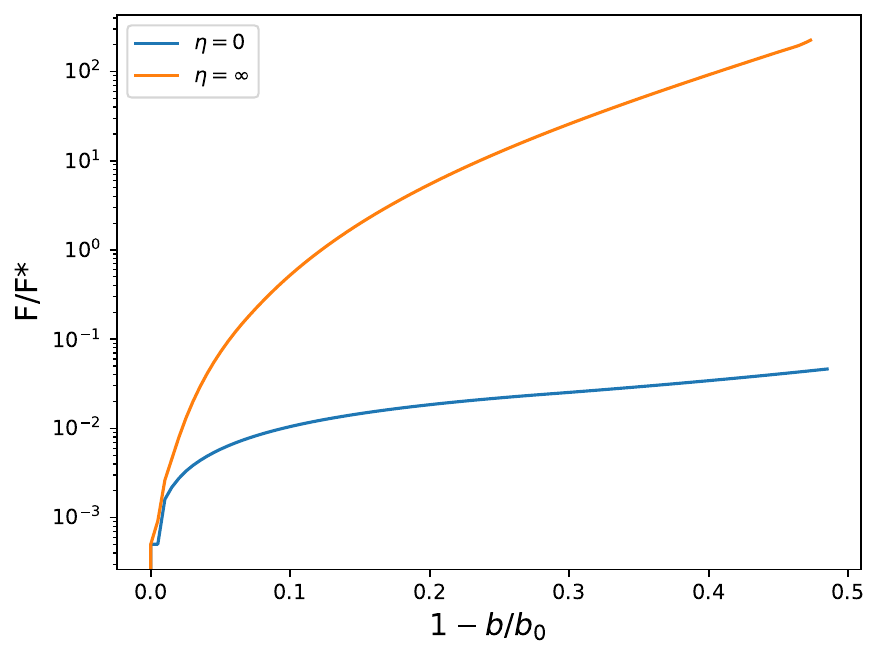}
\caption{The normalized contact force as a function of diametral strain $\varepsilon=1-b/b_0$ for 
for $\eta=0$ (without core stiffness) and for $\eta=\infty$ (with high core stiffness).} 
\label{fig:f-eps}
\end{figure}

\begin{figure}[tbh]
    \centering
    \includegraphics[width=0.42\columnwidth]{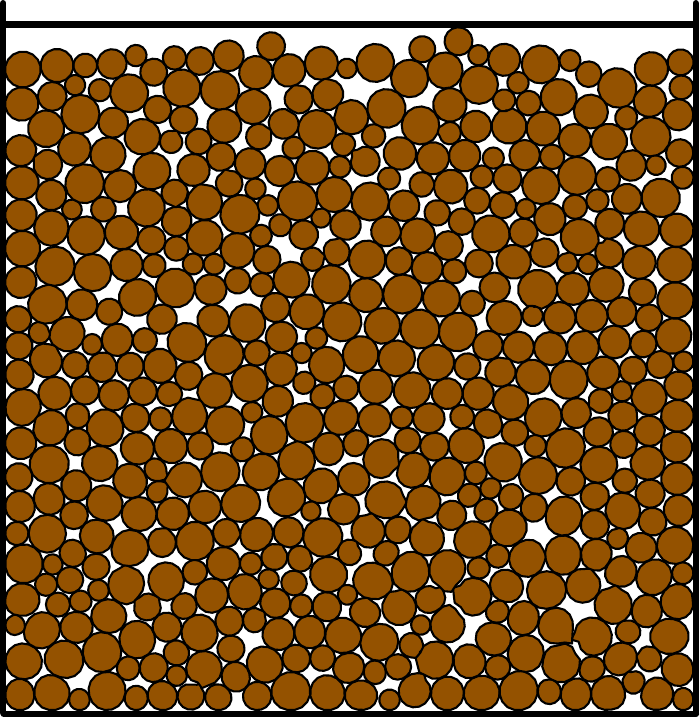}(a)
    \includegraphics[width=0.42\columnwidth]{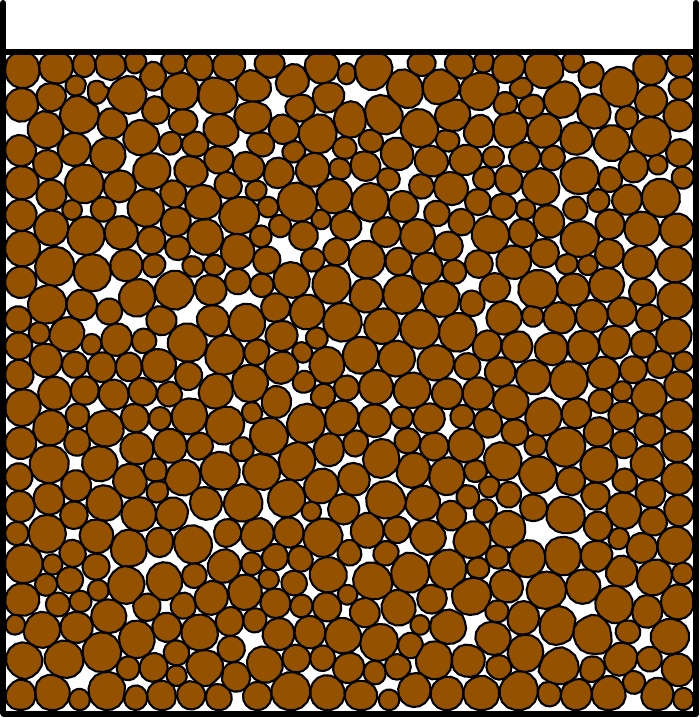}(b)
    \includegraphics[width=0.42\columnwidth]{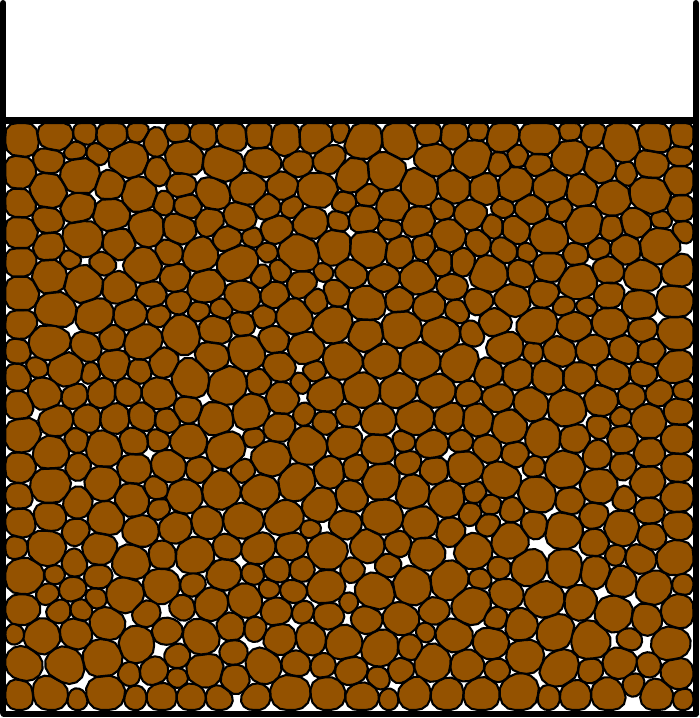}(c)
    \includegraphics[width=0.42\columnwidth]{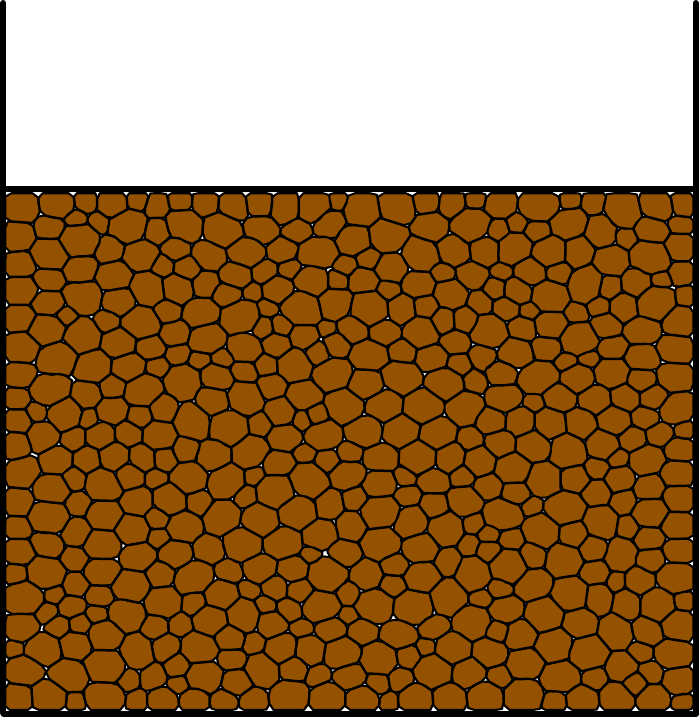}(d)
    \caption{Snapshots of the uniaxial compaction of an assembly of 500 soft core-shell particles.}
    \label{fig:ExampleOedo}
\end{figure}

\begin{figure}[tbh]
    \centering
    \includegraphics[width=0.95\columnwidth]{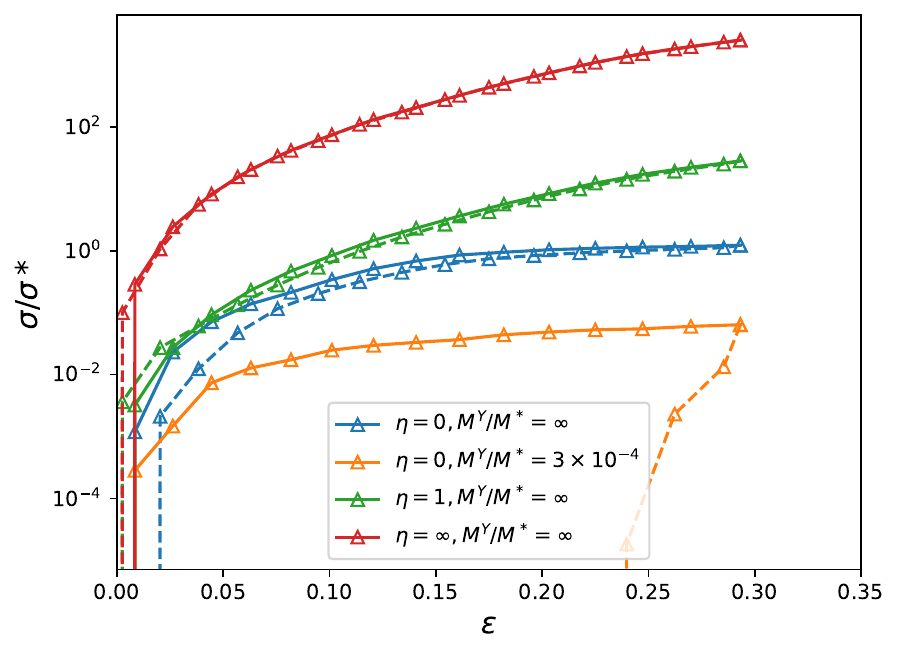}
    \caption{Vertical stress $\sigma$ normalized by the characteristic stress $\sigma^*$ (see 
text) during uniaxial compaction (full lines) and decompaction (dashed lines) 
of an assembly of soft particles for four sets of parameter values: 
1) $\eta=0$ and a high value of the relative plastic 
threshold $M^Y/M^*$, 
2) $\eta=1$ and high $M^Y/M^*$, 
3) $\eta=\infty$ and  high $M^Y/M^*$, and  
4) $\eta=0$ and low value of $M^Y/M^*$.} 
\label{fig:sigma-eps}
\end{figure}

\begin{figure}[h!tb]
\centering
\includegraphics[width=0.8\columnwidth]{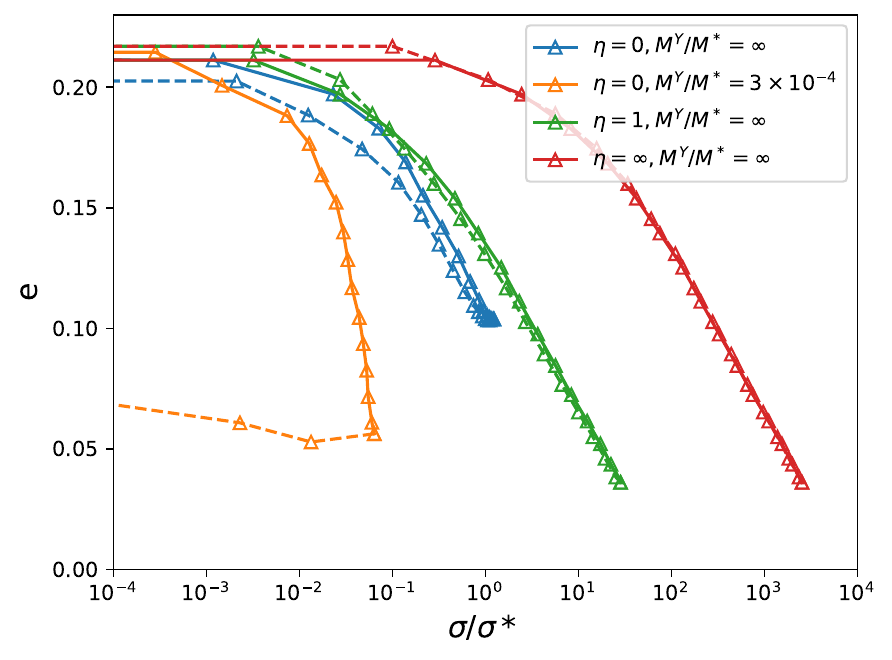}(a)

\includegraphics[width=0.8\columnwidth]{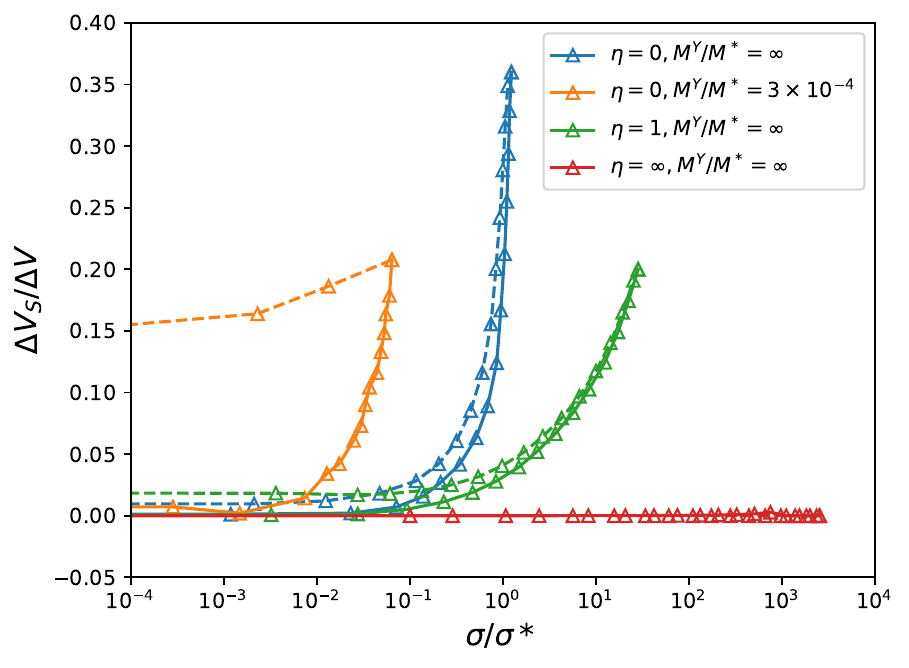}(b)
 \caption{Void ratio $e$ (a) and relative particle volume variation $\Delta V_s/ \Delta V$ (b) 
 as a function of the normalized stress $\sigma/\sigma^*$ for the same 
four parameter sets as in Fig. \ref{fig:sigma-eps}. The solid and dashed lines represent 
the loading and unloading paths, respectively.} 
\label{fig:e-Vs-sigma}
\end{figure}

\begin{figure}[h!tb]
\centering
\includegraphics[width=0.4\columnwidth]{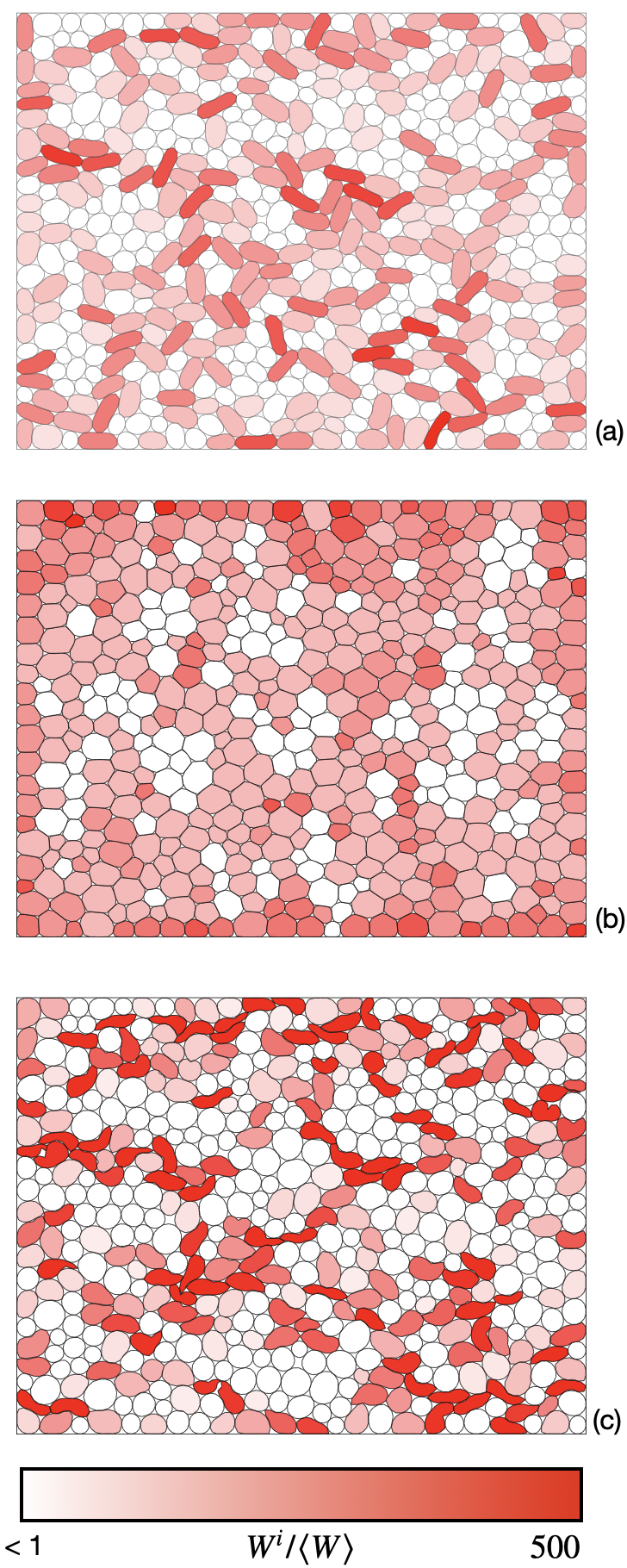}
\caption{Snapshots of the compacted samples for a cumulative 
vertical deformation of $\varepsilon=0.3$ with (a) $\eta=0$ and 
a high value of the relative plastic threshold $M^Y/M^*$ (elastic particles without core stiffness), 
(b) $\eta=\infty$ and  high $M^Y/M^*$ (elastic particles and high core stiffness), 
and  (c) $\eta=0$ and very low value of $M^Y/M^*$ (plastic particles 
without core stiffness). Color levels represent the normalized elastic energies of the particles 
for the energies above the average energy per particle. Particles of elastic energy below 
the average are in white.} 
\label{fig:NRG}
\end{figure}

The total force $ F $ acting on the contact line is the sum of the 
internal force $ F_{in} $ due to pressure $ p $ and the characteristic 
elastic force $ F^* $ at each node, multiplied by the number of 
nodes $ L_c / \ell $:
\begin{equation}
\frac{F}{F^*} = \left( 1 + \frac{ps \ell}{F^*} \right) \frac{L_c}{\ell}.
\label{eqn:L-Fc3}
\end{equation}
The significance of the additional term, compared to Eq. (\ref{eqn:F-Lc}), 
depends on $ p $, which is a function of the volume change. 
Figure \ref{fig:F-Lc3} illustrates $ F $ as a function of $ L_c $ for 
both $ \eta = 0 $ and $ \eta=\infty$. As expected, for a given $ L_c/\ell $, 
the contact force is higher in the $ \eta=\infty$ case, with the difference 
increasing as volume change and contact length increase.

The pressure distribution along the contact line, shown in Fig. \ref{fig:p-x3}, 
reveals that the force plateau extends over nearly the entire contact area, 
with sharp drops at both ends of the plateau. Unlike the plastic case, 
the plateau level rises with diametral compression due to the increasing 
internal pressure. Hence, Eq. (\ref{eqn:L-Fc3}) provides an accurate 
estimate of the force.
If $ p $ remains constant, $ F $ is a linear function of $ L_c $. 
However, if the volume change linearly depends on diametral strain, $ p $ 
will also vary linearly with $ L_c $, causing $ F $ to increase quadratically, 
similar to a solid cylinder of Young's modulus $ E $ and Poisson's ratio $ \nu $. 
For such a solid cylinder, the analytical expression of $ L_c $ 
as a function of $ F $ is given by \cite{Johnson1985}:
\begin{equation}
L_c = \left\{ 2F \frac{R}{\pi s} \frac{1-\nu^2}{E} \right\}^{1/2}.
\label{eqn:LcF} 
\end{equation}

It is important to note that the volume change at small strains 
(below 10\%) is negligible, and the particle behaves like an elastic ring without 
core pressure, with the contact force increasing linearly with contact area. 
This behavior is clearly observed in Fig. \ref{fig:F-Lc3} at low values of $ L_c/\ell $.
In plant cells, the hydrostatic turgor pressure plays a similar role with 
respect to cell walls as core pressure does in our model with respect to 
shell elements. The incompressibility of the liquid prevents significant 
volume change, so the turgor pressure primarily facilitates the faceting of 
cell walls during interactions \cite{Kanahama2023,Ali2023}.

Figure \ref{fig:f-eps}  shows the normalized vertical force $F/F^*$ as a function of 
the diametral strain $\varepsilon = 1-b/b_0$ for $\eta=0$ and $\eta=\infty$. 
In both cases, the force increases rapidly (faster than exponential) 
with $\varepsilon$ at the beginning of 
compression, but asymptotically tends to a a linear form for $\eta=0$ 
and a nearly exponential form for $\eta=\infty$. This indicates that 
the volumetric strain decreases almost exponentially with strain. This decrease is 
small for $\eta=\infty$, as observed in Fig. \ref{fig:V3}, but its effect is amplified 
by the high value of the core stiffness.

\section{Uniaxial compaction of an assembly of soft disks}
\label{sec:compaction}

In this section, we present simulations of the uniaxial compaction of an 
assembly of soft frictionless circular particles. The particles are initially allowed 
to fall into a rectangular box under their own weight. After the sample reaches 
static equilibrium, gravity is removed, and the top wall is moved 
downward at a constant velocity. The simulation involves 500 particles 
with an average radius $ R $ and a standard deviation $ \Delta R/R = 0.15 $.

Figure \ref{fig:ExampleOedo} shows snapshots of the uniaxial compaction 
process up to a cumulative vertical strain $ \varepsilon = \ln(H/H_0) \simeq 0.3 $, 
where $ H $ is the current sample height and $ H_0 $ is the height at 
the first point of contact between the top wall and the top layer of 
particles as seen in Fig. \ref{fig:ExampleOedo}(a).
During compaction, the void ratio $ e $, defined as the ratio of the 
total volume $ V_r $ of the pore space between particles to the total 
volume $ V_s $ of particles, decreases due to three factors: 
1) particle rearrangements, 2) particle volume reduction, and 
3) particle shape change. The extent to which each factor contributes to 
pore reduction depends on the elasto-plastic parameters $ M^Y $ and $ \eta $.

Figure \ref{fig:sigma-eps} shows the vertical stress $ \sigma $ 
as a function of the vertical strain $ \varepsilon$ during both the 
loading (downward movement of the top wall) and unloading 
(upward movement of the top wall) phases. 
The data is presented for four distinct sets of parameter values: 
1) $\eta=0$ and a high value of the relative plastic 
threshold $M^Y/M^*$ (elastic particles without core stiffness), 
2) $\eta=1$ and high $M^Y/M^*$ (elastic particles and low core stiffness), 
3) $\eta=\infty$ and  high $M^Y/M^*$ (elastic particles and high core stiffness), 
and  4) $\eta=0$ and very low value of $M^Y/M^*$ (plastic particles 
without core stiffness). The initial configuration at $ h = h_0 $ is the same for  
all the four simulations. 
The stresses are normalized by the characteristic stress
\begin{equation}
\sigma^* = \frac{F^*}{2Rs}.
\end{equation}  

For purely elastic particles (high $ M^Y/M^* $), $ \sigma/\sigma^* $ 
increases in a strongly nonlinear manner with $ \varepsilon$ during loading. 
Upon unloading, the stress decreases along a path that is close to but 
slightly below the loading path, indicating a residual strain. 
This residual strain suggests that due to particle rearrangements, 
the sample ends up in a slightly more compact state than initially, 
while the particles regain their original circular shape. 
The residual strain is smaller for higher values of $ \eta $. 
This collective granular plasticity  
is typically observed in granular materials composed of rigid particles .
In contrast, for plastic deformable particles (with a very low $ M^Y/M^* $), 
the particles undergo irreversible deformation during loading, 
and $ \sigma/\sigma^* $ decreases  
until a significant residual strain is reached. Here, the plastic behavior 
is primarily attributed to the intrinsic plasticity of the particles rather 
than their rearrangement. The observed increase in $ \sigma/\sigma^* $ 
with increasing $ \eta $ reflects the corresponding rise in particle bulk stiffness.

Figure \ref{fig:e-Vs-sigma}(a) displays the relationship between the 
void ratio $ e $ and the normalized stress $ \sigma/\sigma^* $ for the same 
parameter sets discussed in Fig. \ref{fig:sigma-eps}. 
For a total cumulative compaction $ \varepsilon= 0.3 $, 
the void ratio $ e $ decreases from 0.22 to 0.1 for $ \eta = 0 $ without 
core stiffness, to 0.04 for $ \eta = 0 $ with core stiffness, and 
to 0.06 in the case of plastic rings. These variations in $ e $ correspond 
to changes in particle shapes under stress.
During unloading, $ e $ increases across all elastic cases and 
follows a trajectory close to that during loading. When the vertical strain 
returns to zero, $ e $ remains slightly lower than the initial value $ e_0 $, 
indicating that despite the particles' fully elastic behavior, 
there is a small residual compaction due to particle rearrangements. 
In the case of plastic rings, where $ M^Y/M^* $ is low, the change in 
particle shape is irreversible, leading to a slight increase in $ e $ due to 
elastic restitution. However, $ e $ retains a low value after full unloading, 
attributed to both particle plastic deformation and particle rearrangements.

The void ratio $ e $ decreases as a function of $ \sigma/\sigma^* $ initially 
in a nonlinear manner but, at higher stresses, follows a logarithmic 
function over  two decades:
\begin{equation}
e = e_0 - C \log_{10} \left(\frac{\sigma}{\sigma^*} \right), 
\end{equation} 
where $ C $ represents the compressibility index. 
This compaction law is well documented for clays, 
although its microscopic origins are not fully understood \cite{Mitchell2005}. 
In our simulations, the value of $ C $ is approximately 0.07 for elastic particles, 
regardless of the stiffness ratio $ \eta $. However, for plastic particles, $ C $ 
is higher, reflecting their increased deformability.

The volume-change behavior of granular materials is typically 
described using the void ratio $ e $ or the packing fraction $ \Phi = 1/(1+e) $. 
This variable is sufficient when the total particle volume is conserved  
during compaction or flow. However, when particles undergo volume changes, 
it is necessary to consider also the ratio $ \Delta V_s/ \Delta V $, 
where $ \Delta V_s $ represents the cumulative change in particle 
volume from its initial state, and $ \Delta V $ denotes the cumulative 
volume change of the entire sample. This ratio quantifies the contribution 
of particle volume change to the overall compaction.

Figure \ref{fig:e-Vs-sigma}(b) shows $ \Delta V_s/ \Delta V $ as a 
function of $ \sigma/\sigma^* $. For $ \eta=\infty$, 
$ \Delta V_s $ is negligible. However, as $ \eta $ decreases, $ \Delta V_s/ \Delta V $ 
increases significantly, reaching a peak value of 0.37 for $ \eta = 0 $ 
when $ \sigma \approx \sigma^* $ (corresponding to $ \varepsilon= 0.3 $). 
This means that at this point, particle volume change accounts 
for 37\% of the total sample volume change $ \Delta V $. 
The remainder of the volume change is due to changes in the pore volume 
$ \Delta V_r $, resulting from particle shape alterations and, to a lesser extent, 
particle rearrangements. This high volume change is consistent with 
the significant particle volume change observed during diametral 
compression of elastic rings.

In the case of plastic particles, the volume change contribution at $ \varepsilon= 0.3 $ 
is lower, with $ \Delta V_s/ \Delta V \approx 0.22 $. This is because plasticity 
allows particles to change shape more easily than volume.
Snapshots of the samples at $ \varepsilon= 0.3 $ (just before unloading) 
are shown in Fig. \ref{fig:NRG}, along with color-coded levels indicating the elastic 
energies $ W^i $ of particles $ i $, normalized by the average elastic 
energy per particle $ \langle W \rangle $. In all cases, the elastic energy 
distribution is inhomogeneous and closely related to particle shapes.
For elastic particles without core stiffness (Fig. \ref{fig:NRG}(a)), there is a large 
number of rounded-shaped particles with low energy, alongside elongated, 
stadium-like particles with high energies, often forming locally ordered groups. 
In contrast, elastic particles with high core stiffness (Fig. \ref{fig:NRG}(b)) 
deform into polygonal shapes that efficiently fill space, similar to observations 
in MPM simulations of solid disks \cite{Nezamabadi2015,Nezamabadi2024}. 
This polygonal structure arises 
from the ability of elastic rings with isotropic core stiffness to create 
large surface areas. For plastic particles (Fig. \ref{fig:NRG}(c)), 
many particles maintain an elliptical shape, while others are strongly 
distorted due to their plasticity, which allows them to better conform to 
their local environment.
The efficient filling of pore space in Figs. \ref{fig:NRG}(b) and \ref{fig:NRG}(c), 
with void ratios $ e \approx 0.04 $ and $ e \approx 0.06 $, respectively, 
results from two distinct mechanisms: the creation of flat interfaces 
in Fig. \ref{fig:NRG}(b) and the adaptation of particle shapes to the pore 
space in Fig. \ref{fig:NRG}(c). Notably, the void ratio is also higher in regions with 
elliptical particles as compared with the 
initial configuration of circular particles since even slight deviations from circular or spherical 
shape can significantly increase the void ratio \cite{SaintCyr2012,Tran2024}.

\begin{figure}[h!tb]
    \centering
    \includegraphics[width=0.4\columnwidth]{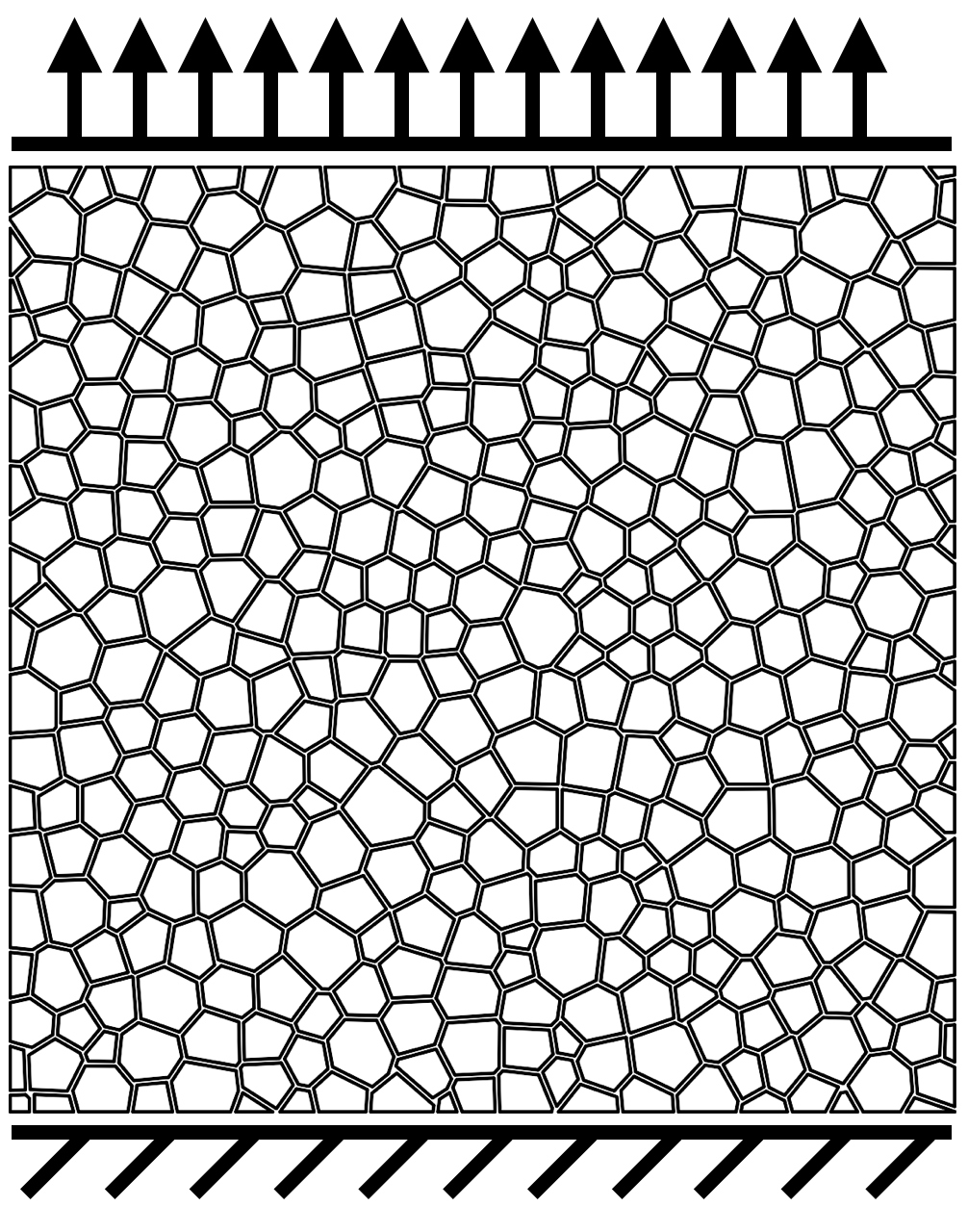}
    \caption{A cohesive cellular tissue subjected to axial extension. The bottom wall is 
    immobile.} 
\label{fig:cellular_scheme}
\end{figure}

\begin{figure}[tbh]
    \centering
    \includegraphics[width=0.45\columnwidth]{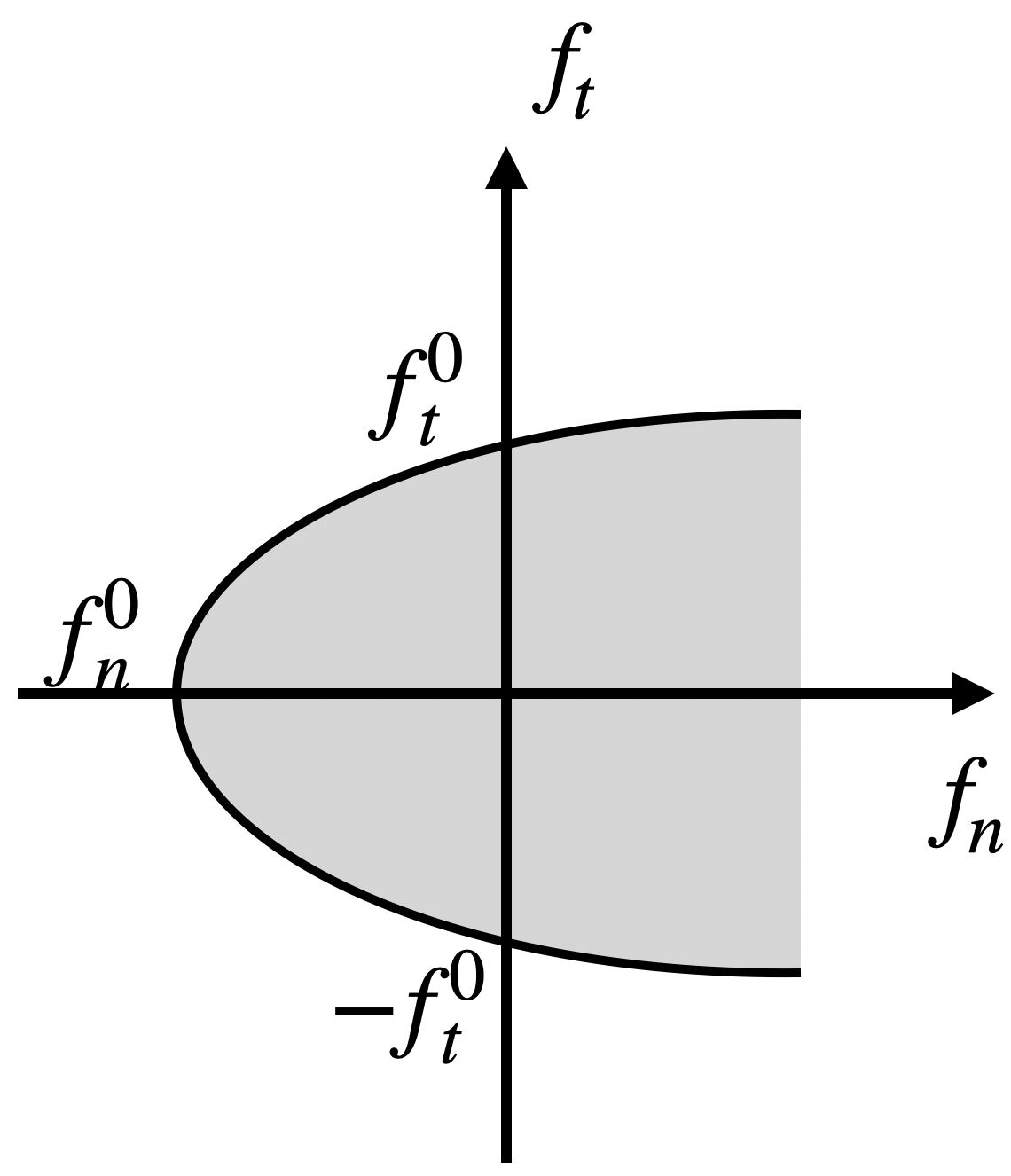}
    \caption{The yield criterion at the interface between two cells; see Eq. (\ref{eqn:zeta}.)} 
\label{fig:yield}
\end{figure}

\section{Fracture of cellular materials}
\label{sec:fracture}

To further demonstrate the applicability of soft particle dynamics, we consider the 
simulation of fracture in cellular tissues subjected to simple tension. 
Figure \ref{fig:cellular_scheme} shows an example of a cellular tissue 
fixed at the bottom and subjected to vertical extension by moving the top wall.
The cellular configuration is created using a Laguerre-Voronoi 
tessellation of a granular packing with small size polydispersity. 
The cells are then transformed into sphero-polygons with a small 
Minkowski radius $ r $.
The tissue's mechanical integrity is maintained by normal and tangential 
cohesive forces between the vertices of adjacent cells. 
For two neighboring vertices at the ends of the interface between 
adjacent cells, the initial distance is set to $ r $. From this reference 
state, we define the normal displacement $ u_n $ and tangential 
displacement $ u_t $, corresponding to normal and tangential forces $ f_n $ 
and $ f_t $, respectively. The normal displacement $ u_n $ is perpendicular 
to the common interface, while the tangential displacement $ u_t $ is parallel to it. 

We assume linear force-displacement relationships for both 
normal and tangential forces, such that $ f_n = k_n u_n $ 
and $ f_t = k_t u_t $, where $ k_n $ and $ k_t $ represent the normal 
and tangential stiffness coefficients of the interface, respectively. 
The sign convention used defines compressive forces and inward displacements as positive.

\begin{figure}[h!tb]
    \centering
    \includegraphics[width=0.35\columnwidth]{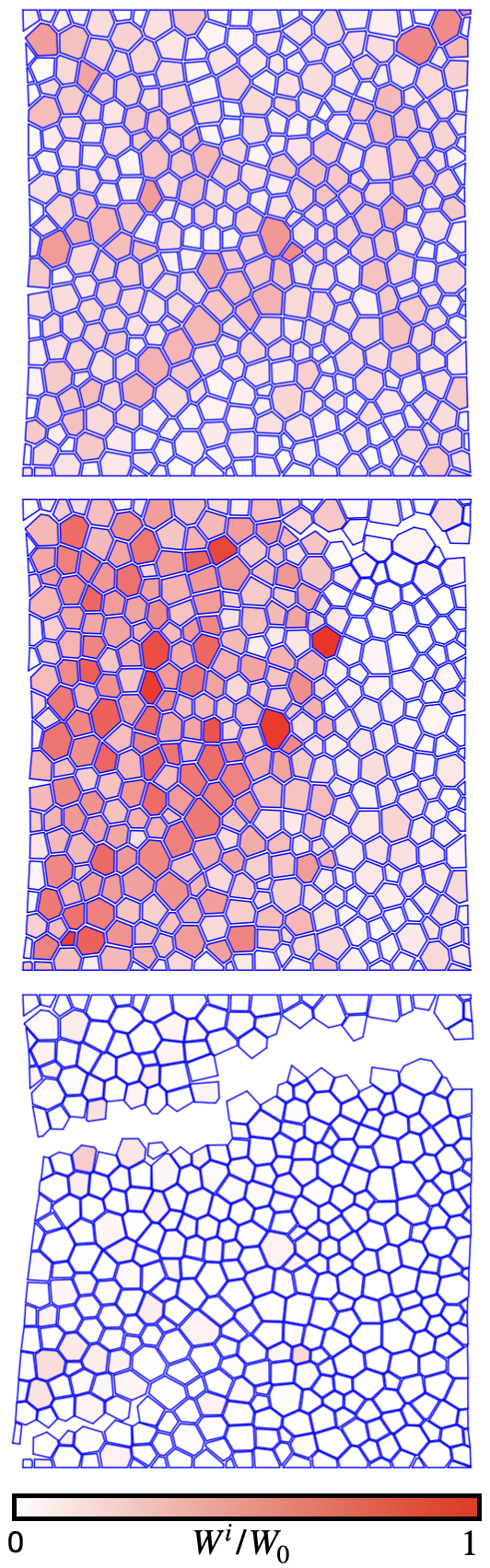}
    \caption{Three snapshots of the tensile loading of a cellular tissue. 
    The color levels are proportional to the elastic energies of the cells normalized by the 
    fracture energy $W_0$ of a single normal cohesive link.} 
 \label{fig:visu-rupture}
 \end{figure}
 
\begin{figure}[tbh]
    \centering
    \includegraphics[width=0.95\columnwidth]{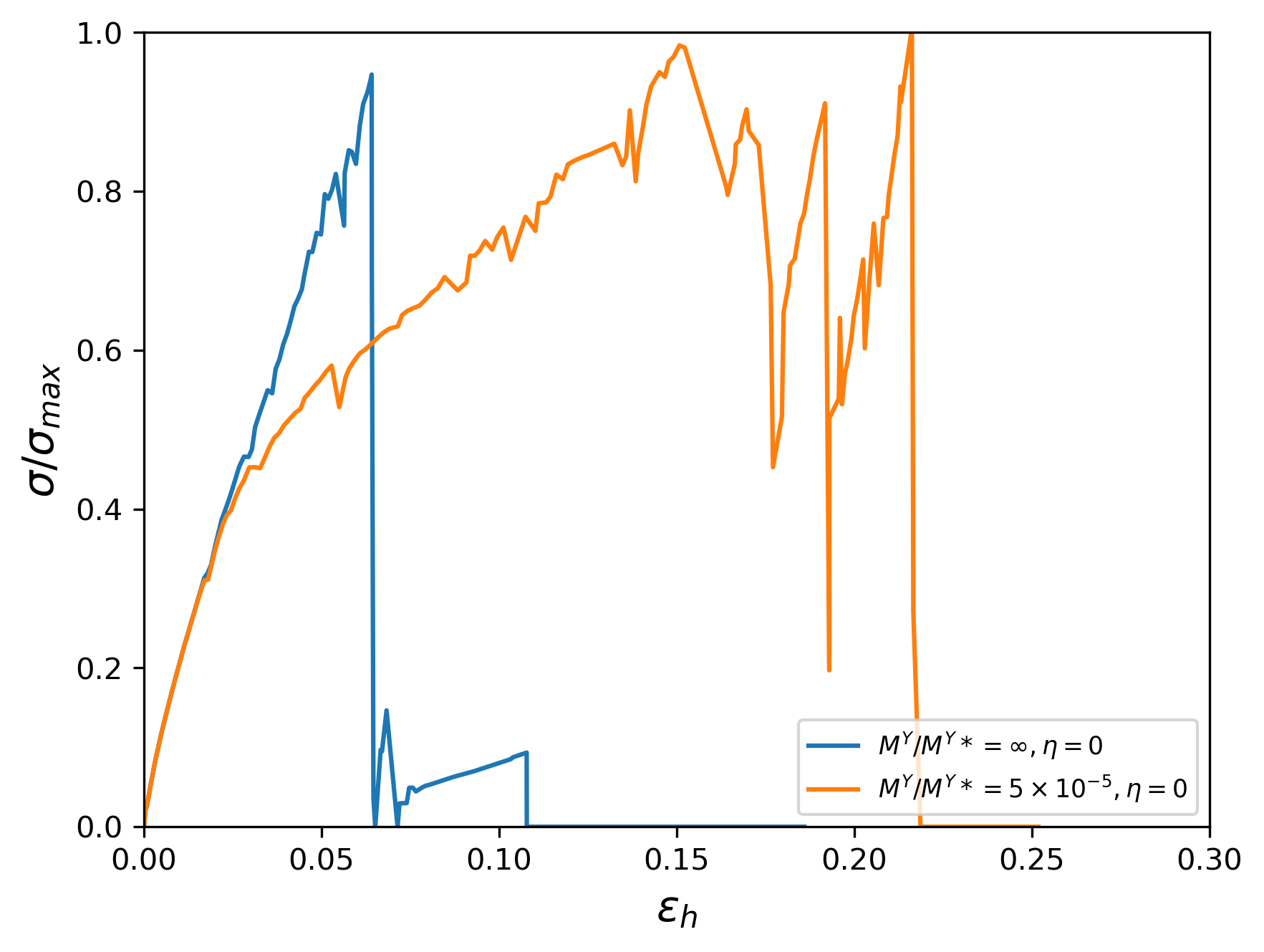}
    \caption{Vertical stress normalized by the peak stress $\sigma_{max}$ versus 
    vertical strain $\varepsilon_h$ for an elastic ring (large value $M^Y/M^*$ and $\eta=0$) and 
    a plastic ring   (small value of $M^Y/M^*$ and $\eta=0$).} 
 \label{fig:sigma-eps}
 \end{figure}

We define a fracture criterion based on a local yield function defined as:
\begin{equation}
\zeta = \frac{f_n}{f_n^0} + \left\vert \frac{f_t}{f_t^0} \right\vert ^\alpha - 1, 
\label{eqn:zeta}
\end{equation}
where $ \alpha $ is a model parameter, $ f_n^0 < 0 $ is the tensile normal 
force threshold in the absence of shear, and $ f_t^0 > 0 $ is the tangential force 
threshold at zero normal force. This criterion is illustrated in Figure \ref{fig:yield} 
for $ \alpha = 2 $. Inside the yield surface ($ \zeta < 0 $), displacements 
follow elastic force laws. A link between two vertices breaks when $ \zeta = 0 $ 
is reached, signifying the onset of fracture. According to this equation, 
mode I fracture (normal displacement) occurs only under tension  
while in compression ($ f_n > 0 $), a link can break only due to shear.

For our simulations, we set $ \alpha = 2 $, as used in previous studies \cite{Delenne2004}. 
When links between cells break, the cohesive connection is lost irreversibly, 
and the interaction changes to a frictional contact with a coefficient of 
friction $ \mu = 0.3 $. We also set $ k_n = k_t = 0.1 k_n $. 
Figure \ref{fig:visu-rupture} shows three snapshots of a tissue 
undergoing tensile loading, along with the elastic energy distributions 
of the cells. The energy values are normalized by the fracture 
energy $ W_0 = (f_n^0)^2 / 2k_n $ of a single normal cohesive bond. 
The first cracks typically appear near the boundaries, but only 
one crack propagates across the tissue, following a rough intercellular path.

\begin{figure}[h!tb]
    \centering
    \includegraphics[width=0.4\columnwidth]{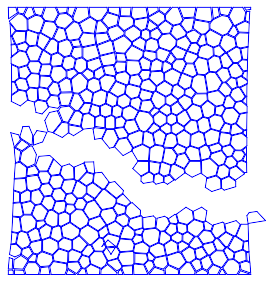}(a)
    \includegraphics[width=0.4\columnwidth]{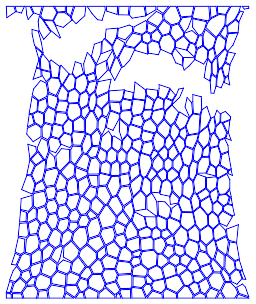}(b)
    \caption{Snapshots of the elastic (a) and plastic (b) cellular tissues 
    fractured under tensile loading.} 
 \label{fig:snap-rupture}
 \end{figure}

In Figure \ref{fig:sigma-eps}, we observe the vertical stress as a function 
of vertical strain for both elastic and plastic particles for $ \eta = 0 $. 
The elastic case exhibits brittle behavior: the stress increases with 
strain until reaching a peak, followed by a sharp drop. Cracking events 
manifest as fluctuations in the stress before fracture. In contrast, the plastic 
case shows initial elastic deformation followed by a prolonged plastic 
phase, characterized by stress drops of increasing magnitude. 
Snapshots of the fracture path (Figure \ref{fig:snap-rupture}) 
reveal elongated cells and an irregular fracture path in the plastic case at failure.

\section{Conclusion}
\label{sec:conclusion}

In this paper, we introduced a Soft Particle Dynamics (SPD) model for simulating the 
dynamics of collections of soft core-shell particles such as granular 
materials composed of deformable particles 
and biological tissues. To accommodate large particle 
deformations, the particle surface is represented by a shell of mass points 
interacting through linear and angular stiffness, as well as a core stiffness. 
The model incorporates also frictional contact interactions between particles, 
shape plasticity, cohesive links, and fracture mechanics. 
Unlike continuum approaches, our SPD approach extends the Discrete Element 
Method (DEM) to include shape degrees of freedom. 

We showed that the model parameters can be mapped to 
material properties such as the elastic and plastic moduli of a 
beam or a circular particle. For verification and validation, we compared the elastic 
deformation of beams under point loads and the compression of particles 
between two walls with theoretical or semi-theoretical predictions. 
These behaviors were analyzed using dimensionless variables 
to understand the influence of parameters such as core stiffness 
and shell plastic threshold on the particle deformation. 

When applied to the uniaxial compaction of a collection of deformable circular 
particles, our numerical method effectively captures different mechanisms 
underlying the compaction process based on the values of model parameters. 
The particles adapt their shapes to fill pore spaces through elongation and  
flat edge-to-edge contacts, or by conforming to the shape 
of the pore space. The compaction process was observed to follow 
a logarithmic consolidation law, similar to that seen in clays. 
We also applied the model to cellular tissues by simulating 
tensile fracture in cellular tissues modeled as bonded polygonal cells.

Although the two-dimensional SPD, as formulated in this paper, 
does not precisely replicate the mechanical 
behavior of solid cylindrical particles—such as the force-surface area 
relationship in diametral compression—due to the degrees of freedom 
being limited to the particle periphery, it closely approximates the behavior 
of solid particles when high core stiffness is present. This leads to minimal 
volume change and the development of large contact areas with walls. 
The method is also particularly well-suited for simulating plant tissues and vesicles, 
with the control of bulk stiffness, internal pressure, or particle volume. 
Similar to the DEM, the computational cost of SPD simulations depends 
on the total number of particles, as well as the number of nodes per 
particle, which determines the accuracy of particle shape representation 
and its deformation.   
Extending the SPD method to three dimensions is straightforward: 
mass points are distributed across each particle surface, with 
their interactions parameterized using linear elements connecting 
these vertices. Contact detection between particles is then based 
on the distances between vertices of particles and the 
surface elements of other particles. As in 2D, the contact 
forces are mapped to the neighboring mass points. 

The SPD model offers significant potential for efficient numerical simulations of 
deformable particle systems. We are currently undertaking an extensive 
parametric study on the compaction of soft granular materials, 
exploring a wide range of model parameters. The findings from this study 
will be detailed in an upcoming paper. While this paper 
primarily addressed examples of nearly homogeneous cellular structures, 
we have also conducted an in-depth investigation into the fracture behavior 
of disordered cellular tissues, with results to be presented in future work.

\section*{Acknowledgment}
We acknowledge financial support from INRAE (TRANSFORM Department)
and University of Montpellier for funding this study.

%
%

 \bibliographystyle{elsarticle-num} 
 \bibliography{references.bib}





\end{document}